 \documentclass[final,authoryear,3p,11pt]{elsarticle}

\usepackage{graphicx}
\usepackage{epstopdf}
\usepackage{booktabs}
\usepackage{natbib}
\usepackage{fancyhdr}
\usepackage{amsmath}
\usepackage{subfig}                     %more grapgh in one figure
\usepackage{amsfonts}
\usepackage{multirow}
\usepackage{lscape}
\usepackage{rotating}
\usepackage{color}
\usepackage{hyperref}
\usepackage{color,soul}
\usepackage{textcomp}
\sethlcolor{green}
\usepackage{ifthen}
\usepackage[latin1]{inputenc}
\usepackage{titletoc}
\usepackage{eso-pic}
\definecolor{123}{rgb}{.9,.9,.9}
\hypersetup{
colorlinks=false,
citecolor=blue,
linkbordercolor={1 1 1}, % {.69 .77 .87} % {1 1 1}
citebordercolor={1 1 1},
urlbordercolor={1 1 1}
}

\usepackage{amssymb}
 \usepackage{amsthm}
 \usepackage{xcolor}

\makeatletter\def\ps@pprintTitle{%
 \let\@oddhead\@empty
 \let\@evenhead\@empty
 \let\@oddfoot\@empty
 \let\@evenfoot\@oddfoot
}
\makeatother

\begin{document}

\begin{frontmatter}

%%% Title, authors and addresses
%
%%% use the tnoteref command within \title for footnotes;
%%% use the tnotetext command for the associated footnote;
%%% use the fnref command within \author or \address for footnotes;
%%% use the fntext command for the associated footnote;
%%% use the corref command within \author for corresponding author footnotes;
%%% use the cortext command for the associated footnote;
%%% use the ead command for the email address,
%%% and the form \ead[url] for the home page:
%%%
%%% \title{Title\tnoteref{label1}}
%%% \tnotetext[label1]{}
%%% \author{Name\corref{cor1}\fnref{label2}}
%%% \ead{email address}
%%% \ead[url]{home page}
%%% \fntext[label2]{}
%%% \cortext[cor1]{}
%%% \address{Address\fnref{label3}}
%%% \fntext[label3]{}
%
\title{Behavioural breaks in the heterogeneous agent model: the impact of herding, overconfidence, and market sentiment.\tnoteref{label1} }

%\title{Behavioural Breaks in the Heterogeneous Agent Model \tnoteref{label1} }
%\title{Herding, overconfidence and market sentiment in the Heterogeneous Agent Model \tnoteref{label1} }

%
%%% use optional labels to link authors explicitly to addresses:
%%% \author[label1,label2]{<author name>}
%%% \address[label1]{<address>}
%%% \address[label2]{<address>}
%
\author[ies,utia]{Jiri Kukacka} \ead{jiri.kukacka@gmail.com}
\author[ies,utia]{Jozef Barunik\corref{cor2}} \ead{barunik@utia.cas.cz}
\cortext[cor2]{Corresponding author}
\address[ies]{Institute of Economic Studies, Faculty of Social Sciences, Charles University, Opletalova 21, 110 00, Prague, Czech Republic}
\address[utia]{Institute of Information Theory and Automation, Academy of Sciences of the Czech Republic, Pod Vodarenskou vezi 4, 182 00, Prague, Czech Republic}

\tnotetext[label1]{We are grateful to Lukas Vacha for many useful discussions. The support from the Czech Science Foundation under the 402/09/0965 project and Grant Agency of the Charles University under the 588912 project is gratefully acknowledged. This is a preliminary draft version compiled \emph{\today}. Please do not use, cite or copy the text without permissions from authors.}

\begin{abstract}
The main aim of this work is to incorporate selected findings from behavioural finance into a Heterogeneous Agent Model using the \citet{BrHo1998} framework. Behavioural patterns are injected into an asset pricing framework through the so-called `Break Point Date', which allows us to examine their direct impact. In particular, we analyse the dynamics of the model around the behavioural break. Price behaviour of 30 Dow Jones Industrial Average constituents covering five particularly turbulent U.S. stock market periods reveals interesting pattern in this aspect. To replicate it, we apply numerical analysis using the Heterogeneous Agent Model extended with the selected findings from behavioural finance: herding, overconfidence, and market sentiment. We show that these behavioural breaks can be well modelled via the Heterogeneous Agent Model framework and they extend the original model considerably. Various modifications lead to significantly different results and model with behavioural breaks is also able to partially replicate price behaviour found in the data during turbulent stock market periods. 
\end{abstract}
%\begin{keyword}
%heterogeneous agent model \sep behavioural finance \sep herding \sep overconfidence \sep market sentiment \sep stock market crash
%\end{keyword}
\end{frontmatter}

%\textit{JEL: C1, C61, D84, G01, G12} \\

%\tableofcontents

%\newpage

\section{Introduction}

The representative agent approach and the Efficient Market Hypothesis \citep{Fama1970} together with the Rational Expectations Hypothesis (\citealp{Muth1961}, \citealp{Lucas1972}), which have dominated the field in the past, are being replaced by more realistic agent based computational approaches in recent literature. These movements in economic thought are reflected in a subset of agent based models, referred to as Heterogeneous Agent Models (HAM henceforth), abandoning agents' full rationality. Agents do not become irrational but `boundedly rational' (\citealp{Simon1955,Simon1957}; \citealp{Sargent1993}), they posses heterogeneous expectations, use simple forecasting rules to predict future development of market prices and the accuracy of their decisions is evaluated retroactively. Based on the simple profitability analysis, agents switch between several trading strategies --- the parallel with the human learning process is more than apparent. Market fractions thus co-evolve over time and interactions between agents endogenously influence market prices which are no longer driven by exogenous news only.

Behavioural finance can be viewed as another answer to unrealistic assumptions of the Efficient Market Hypothesis. It suggests to employ the insights from behavioural sciences such as psychology and sociology into financial market models dating initial work back to the 1970s \citep{KaTv1974,KaTv1979}. Just as HAM do, behavioural finance also builds on the bounded rationality and argues that some phenomena observed in the financial world can be better explained using models with agents which are not fully rational. Some authors even suggest the \textit{``behavioural origin of the stylized facts of financial returns''}, and of the \textit{``statistical regularities of the data"} \citep[pp. 19 \& 39]{AlLuWa2005}. These psychological finding may have significant impacts on the theory of stock trading as they directly violate the Efficient Market Hypothesis \citep{Shiller2003}.

Both approaches --- HAM and behavioural finance --- complement one another and could be used together as HAM framework could serve as a useful theoretical tool for verification of findings from behavioural finance. \citet[pg. 41]{Lebaron2005} argues that \textit{``agent-based technologies are well suited for testing behavioural theories''} and anticipates that \textit{``the connections between agent-based approaches and behavioural approaches will probably become more intertwined as both fields progress''}. Moreover, many studies have highlighted different behavioural patterns as an optimal way of motivating the underlying HAM assumptions of strategy switching and heterogeneous beliefs. \citet{BaTh2003} and \citet{ScXi2004} mention overconfidence, \citet{DeGr2006} and \citet{BoHoMa2007} suggest market sentiment, and \citet{Chang2007} and \citet{ChGaLePa2003} put stress on herding behaviour.

Being an interdisciplinary research on the edge of economics and other fields such as psychology, sociology, and especially physics, interacting agents attracted researchers from many fields. \citet{Kaizojietal2002} innovate a spin model motivated by the Ising model \citep{Ising1925,Onsager1944} and apply it to interpret the magnetization in terms of financial markets and to study the mechanisms creating bubbles and crashes. \citet{Westerhoff2004} discusses the role of emotions such as greed and fear in the determination of stock prices. The work of \citet{Dong2007} extends an interacting herding model by the clustering tendency of agents. An innovative approach introduced in \citet{Shimokawaetal2007} studies the loss-aversion features embedded into an agent-based model. Authors show many consistencies with stylized facts and `puzzles' in financial markets and demonstrate that a rise of loss-aversion amplifies the price distortions. \citet{LaSe2008} propose an original treshold approach to the heterogeneous agent modeling within which the strategy of an agent is defined by a pair of moving thresholds around the current price. This method allows to include various types of agent motivations and behaviours in a consistent manner. \citet{Liuetal2008} consider an effect of the HAM market microstructure, clearing house frequency, and behavioral assumptions on the differences between high-frequency returns and daily returns.  \citet{Kaltwasser2010} presents another simple HAM, where traders have different beliefs about the fundamental rate. The model includes only fundamentalists and even in the absence of trend followers, cyclical fluctuations of the exchange rate can emerge. The most recently, \citet{Diksetal2013} study the effect of memory in the framework of the simple HAM with agents switching between costly innovation and cheap imitation strategies. Authors highlight the fact, that although memory is commonly acknowledged to reduce the variance of prices and quantitatively stabilize the model, it can also have a qualitatively destabilizing effect. We leave another interesting results of this interdisciplinary research \citep{Abdullah2003,Ferreiraetal2005,SaGa2007,Schutzetal2009,Zhuetal2009,Biondietal2012} for the reader's inquisitiveness.
 
The central idea of our work is therefore to take advantage of both approaches and to interconnect particular findings from behavioural finance with heterogeneous expectations in an asset pricing framework in order to study resulting price dynamics. By doing so, we also investigate whether current HAM methodology can be reasonably extended by applying findings from the field of behavioural finance. Or conversely, whether HAM can serve as a tool for theoretical verification of these findings.
%Financial crises and stock market crashes can be widely considered as periods when investors' rationality is restrained and where behavioural patterns are likely to emerge, strengthen and often play the dominant role.  Hence, there is strong rationale to advance current literature through an empirical verification of HAM abilities and explanatory power of behavioural finance, using data covering these turbulent periods. 
Considering HAM methodology, we follow the \citet{BrHo1998} model approach and its extensions. From the plethora of well documented behavioural biases we examine the impact of herding, overconfidence, and market sentiment as these are generally supposed to have a strong impact on traders' behaviour not only during turbulent periods. Standard statistical tools of data analysis together with computational simulations are employed. Specifically, we aim at answering the question if selected findings from behavioural finance can be well modelled via the HAM, extend it and finally replicate price behaviour of real world market data.

For this purpose, we collect a unique dataset of 30 Dow Jones Industrial Average constituents covering five particularly turbulent stock market periods. The sample we consider starts with Black Monday 1987, the largest one-day stock market drop in the history, and terminates with the Lehman Brothers Holdings bankruptcy in 2008, one of the milestones of the recent financial crisis of 2007--2010. We aim at studying the price behaviour around the crash days. In the second part of the work, we develop a simulation based framework where we inject a selected behavioural findings into the HAM and study the behaviour of simulated prices around this break point.

%Thus our main aim is to incorporate selected findings from behavioural finance into a HAM using the the \citet{BrHo1998} framework. In particular, we analyse the dynamics of the model around the so-called `Break Point Date' (BPD henceforth) where behavioural elements are injected into the system and compare it to our empirical benchmark sample based on stock market data surrounding five particularly turbulent stock market periods. Behavioural patterns are thus embedded into an asset pricing framework, which allows to examine their direct impact. The simplicity of this approach also enables us to keep the impact of behavioural modifications as clear as possible. The occurrence of behavioural patterns at the BPD is obviously not perceived as `out of the blue' because the behaviour of market participants is subject to many biases at all times. We interpret it rather as a `considerable amplification' of their impact during market crashes and crisis periods.

The work is structured as follows. After the introduction, we motivate a selected findings from behavioural finance we would like to study and we offer a description of the \citet{BrHo1998} heterogeneous model framework. Next, we study a dataset of different financial crises periods and we find similarities in changes before and after the market crash. Finally, we pay our attention to the numerical analysis and simulation techniques used to study behavioural changes in the HAM.

\section{Selected findings from behavioural finance}

From the plethora of biases, irregularities or seemingly irrational behavioural patterns studied within the field of behavioural finance we focus on three particular findings and extend the model in the following behavioural directions:

\begin{enumerate}
  \item Herding;
  \item Overconfidence;
  \item Market sentiment.
\end{enumerate}

There are several good reasons why we should place special focus on these behavioural biases. First, they are robust and well documented in many studies. Second, they are generally supposed to have strong impact on traders' behaviour over the long run, not only during turbulent periods. Third, all three phenomena can be well integrated into the \citet{BrHo1998} model framework which is rather compact and does not otherwise allow for major modifications without deviating from its overall structure. Let us briefly motivate each of the biases.

%Further, as it goes far beyond the scope of this work, we neither describe nor discuss here the plethora of remaining behavioural biases. To gain more information about this fascinating field of finance, please, let yourself be inspired in \autoref{sec:bflr}.

%***********************

\subsection{Herding}
\label{sec:herding}

Herding (or herd behaviour) describes a situation when many people make similar decisions based on a specific piece of information while ignoring other highly relevant facts. \citet{Keynes1936} comments on herding tendencies when he describes the stock market as a `beauty contest'. For a financial market example, if stock prices go up, it is likely to attract public attention and allows for irrational enthusiasm that can develop into a market bubble in the end. High expectations of future prices are the reason for current high prices and vice versa, no matter that there is likely to be no real merit behind these expectations. As in a herd, people `follow the crowd' in terms of both expectations and real investment decisions. Momentum trading or positive feedback trading can serve as good examples.

However, herding might be far from being irrational. On the individual level, imitating of actions of others --- e.g. market leaders --- might be an extremely cheap, easy, and effective way of learning and decision-making. \citet{Chang2007} for instance understands herding as an evolutionary adaptation, which developed naturally as a cost-effective way of processing information. Moreover, for traders professing market psychology, moving against the herd may present an attractive investment strategy. On the other hand, for economy as a whole, herding is likely to decrease market efficiency and can even have disastrous impact.

Generally, people do not like uncertainty and behave according to observed patterns even if it is often hard to find any objective reasoning supporting such a strategy \citep{Shiller2003}. People are also highly influenced by their environment and the world of finance is no exception. For animals, safety is one of fundamental reasons why they herd and for investors, professional money managers, or analysts the same holds in many situations \citep{DeTh1995}. These interesting explanations are given by \citet[pg. 750]{DiWe2005}: \emph{``being too different from the rest can be risky and might jeopardize career perspectives or reputation''} or \emph{``younger analysts forecast closer to the average forecast''}, as \emph{``they are more likely to be terminated when they deviate from the consensus''}.

Herding is sometimes considered as an opposite tendency to overconfidence regarding information efficiency. \citet[pg. 326]{BeWe2001} for example argues that thanks to overconfident individuals, information \emph{``that would be lost if rational individuals instead just followed the herd''} is preserved. For brevity reasons, we do not offer a complete literature review on the topic of herding but do refer to several key contributions. For concerned readers, \citet{Hirshleifer2001}, \citet{DiWe2005}, \citet{AlLuWa2005}, or \citet{Hommes2006} are likely to serve as a good sources of information. Modelling approaches to herding are offered by \citet{Kirman1991,Kirman1993}, who studies herding behaviour in ant colonies and might be considered as a `promoter' of modelling of this mechanism. A HAM discussing herding behaviour is introduced by \citet{ChHe2002}, who reveal a tendency to herd when a particular strategy becomes significantly profitable. \citet{ChGaLePa2003} confirm  the potential of imitating to be a rational strategy. \citet{DiWe2005} suggest that herding might increase market volatility.

%***********************

\subsection{Overconfidence}
\label{sec:ovce}

Many psychological studies indicate that people are generally overconfident. In a nutshell, overconfidence is a consistent tendency to overestimate own's skills and the accuracy of one's judgments. People often believe in their own superior knowledge and put much more weight on private information --- especially if they are personally involved in gathering and assessing data --- than on public signals --- particularly when these are ambiguous. People also poorly estimate probabilities of future events and are too optimistic about future success --- especially when it comes to challenging tasks. \citet[pg. 389]{DeTh1995} even comment on overconfidence as \textit{``perhaps the most robust finding in the psychology of judgment''}.

The most famous example to illustrate overconfidence concerns driving abilities. From a sample of 81 U.S. students 82\% believed they were in the top 30\% of drivers in terms of driving safety and almost 93\% found themselves above average in terms of driving skills \citep{Svenson1981}. Another interesting fact is that men have been found to be more overconfident than women and experts more overconfident than laymen \citep{BaTh2003}.

Overconfidence is an extremely relevant topic for financial markets. Investors overconfident about their trading abilities are prone to pursue excessive trading (\citealp{Odean1998,Odean1999}; \citealp{BaOd2000,BaOd2002}), hold under-diversified portfolios \citep{GoKu2008}, or underestimate risk \citep{DeBondt1998}. All these factors are likely to implicate higher transaction costs hand-in-hand with lower returns \citep{BaOd2000}. On the other hand, one of positive implications of overconfident behaviour might be the reduced tendency to herd \citep{BeWe2001}.

Regarding further literature, we mention a few typical theoretical approaches to overconfidence modelling. Interested readers are, however, referred to \citet{DaHiSu1998}, \citet{Hirshleifer2001}, or \citet{BaTh2003} to gain more information about the topic and other modelling approaches. \citet{DaHiSu1998} present a model of securities over- and under-reaction based on overconfidence which is defined as the overestimation of private signals precision.\footnote{Which is consistent with our description of overconfidence but in sharp contrast with the model of investor sentiment by \citet{BaScVi1998} mentioned in \autoref{sec:ms} in which all information is public but subject to misinterpretation. However, the results of both models regarding the effect on market prices are comparable.} In their work they reveal a tendency of overconfidence to impact market prices and increase market volatility. In many studies, overconfidence is modelled as \textit{``overestimation of the precision of one's information''} \citep[pg. 15]{ScXi2004}. In another way of looking at it, \citet{BaTh2003} for instance suggest that overconfidence can be modelled as an underestimation of variance.

%***********************

\subsection{Market sentiment}
\label{sec:ms}

Last finding from behavioural finance which we use is market sentiment. Defined broadly, market sentiment refers to exaggeratedly pessimistic or optimistic and wishful beliefs about future market development, stock cash flows, and investment risks which are not fully justified by information at hand. To the best of our knowledge, market sentiment seems to be one of the most powerful driving forces on the stock market ---  as  early  as \citeyear{Keynes1936} \citeauthor{Keynes1936} highlighted the role of sentiment as one of major determinants of investment decisions. This is particularly true during market crashes --- \citet[pg. 1934]{DeVeZw2009a} for instance point out that \emph{``there is a clear shift in sentiment during extreme events''} and a study by \citet{Shiller2000} shows that investor sentiment in terms of bubble expectation and investor confidence vary significantly through time. Market sentiment causes irrational shifts of aggregate demand or supply as the behaviour affected by market sentiment is correlated among traders \citep{ScSu1990}. These shifts might be triggered for various reasons. Investors might follow `market gurus' or expert advice, use the same pricing models or sources of information (rating agencies), react rashly to signals they do not fully understand, or just follow the crowd.

The academic literature concerning market sentiment from various points of view is rather extensive. Here, we briefly refer to several interesting works. An article by \citet{BaScVi1998} introduces a model of investor sentiment which is developed to reflect psychological as well as empirical evidence of overreaction and under-reaction of stock prices. A study by \citet{BaRuWu2005} briefly suggests several approaches to sentiment and overconfidence modelling. \citet{BaWu2007} develop the \emph{Sentiment Index} --- a methodology of measuring investor sentiment --- and describe in detail which empirical proxies to employ for its creation. \citet{BoHoMa2007} propose a simple theoretical framework of measuring the average market sentiment within the HAM framework. Finally, in their article, \citet{VaBaVo2009} consider the simple form of market sentiment in the HAM framework.

%***************

\section{Framework for heterogeneous agents model}
\label{sec:model}

Our modeling framework is within the \citet{BrHo1998} HAM. The model is a financial market application of the adaptive belief system --- the endogenous, evolutionary selection of heterogeneous expectation rules following the framework of \citet{Lucas1978} and proposed in \citet{BrHo1997,BrHo1998}. We consider an asset pricing model with one risk free and one risky asset. The dynamics of the wealth is as follows:
\begin{equation}\label{eq:a}
W_{t+1}=RW_{t}+(p_{t+1}+y_{t+1}-Rp_t)z_t,
\end{equation}
where $W_{t+1}$ stands for the total wealth at time $t+1$, $p_t$ denotes the ex-dividend price per share of the risky asset at time $t$, and \{$y_t$\} denotes its stochastic dividend process. The risk-free asset is perfectly elastically supplied at constant gross interest rate $R=1+r$, where $r$ is the interest rate. Finally, $z_t$ denotes the number of shares of the risky asset purchased at time $t$. The type of utility function considered is essential for each economic model and determines its nature and dynamics. The utility for each\footnote{This is a crucial assumption without which the original model of \citet{BrHo1998} looses one of its greatest advantages of analytical tractability.} investor (trader or agent alternatively) $h$ is given by $U(W)=-exp(-aW)$, where $a>0$ denotes the risk aversion, which is assumed to be equal for all investors.\footnote{The generalized version of the model with the aim to study the model behaviour after relaxing a number of assumptions --- especially homogeneous risk aversion --- has been proposed by \citet{ChHe2002b}. Authors allow agents to have different risk attitudes by generalizing \autoref{eq:b} and letting the risk aversion coefficient $a_h$ differ among particular traders. The paper is focused mainly on a study of two-belief systems. Typically, fundamentalists are expected to be more risk averse than chartists and thus the relative risk ratio $a_{cf}=\frac{a_{chart.}}{a_{fund.}}<1$. Authors offer an exuberant analysis of many specific setting combinations and conclude that relaxing some assumptions of the original \citet{BrHo1998} model leads to a markedly enriched system with some significant differences (e.g. stability of the model equilibrium might depend directly on $a_{cf}$; decreasing $a_{cf}$ may trigger chaotic fluctuations around the fundamental price, i.e. when fundamentalists are more risk averse, market becomes more chaotic; or that adding noise has a small effect when $a_{cf}$ is large, but the opposite is true when $a_{cf}$ is small). On the other hand, many of the original results are robust enough with regard to suggested generalizations.} For determining the market prices in this model, the Walrasian auction scenario is assumed. I.e. the market clearing price $p_t$ is defined as the price that makes demand for the risky asset equal to supply at each trading period $t$ %This in fact means the price $p_t$ at time $t$ if derived employing information from time $t-1$ and the expected utility for time $t+1$.%
and investors are `price takers'. The detailed description of the price formation mechanism is offered further in this section and finally summarized by \autoref{eq:n} and \autoref{eq:new}.

Let $E_t$,~$V_t$ denote the conditional expectation and conditional variance operators, respectively, based on a publicly available information set consisting of past prices and dividends, i.e. on the information set $\mathcal{F}_t=\{p_t,p_{t-1},\dots;y_t,y_{t-1},\dots\}$. Let $E_{h,t},~V_{h,t}$ denote the beliefs of investor type $h$ (trader type $h$ alternatively) about the conditional expectation and conditional variance. For analytical tractability, beliefs about the conditional variance of excess returns are assumed to be constant and the same for all investor types, i.e. $V_{h,t}(p_{t+1}+y_{t+1}-Rp_t)=\sigma^2$. Thus the conditional variance of total wealth $V_{h,t}(W_{t+1})=z^2_t \sigma^2$.

Each investor is assumed to be a myopic\footnote{To be `myopic' means to have a lack of long run perspective in planning. Roughly speaking, it is the opposite expression to `intertemporal' in economic modelling.} mean variance maximizer, so for each investor $h$ the demand for the risky asset $z_{h,t}$ is the solution of:
\begin{equation}\label{eq:b}
\max_{z_t}\left\{E_{h,t}[W_{t+1}]- \frac{a}{2} V_{h,t} [W_{t+1}]\right\}.
\end{equation}
Thus
\begin{equation}\label{eq:c}
E_{h,t}[p_{t+1}+y_{t+1}-Rp_t]-a \sigma^2 z_{h,t}=0,
\end{equation}
\begin{equation}\label{eq:d}
z_{h,t}=\frac{E_{h,t}[p_{t+1}+y_{t+1}-Rp_t]}{a \sigma^2}.
\end{equation}
Let $n_{h,t}$ be the fraction of investors of type $h$ at time $t$ and its sum is one, i.e. $\sum_{h=1}^{H} n_{h,t}=1$. Let $z_{s,t}$ be the overall supply of outside risky shares. The Walrasian temporary marker equilibrium for demand and supply of the risky asset then yields:
\begin{equation}\label{eq:e}
\sum_{h=1}^{H} n_{h,t} z_{h,t}=\sum_{h=1}^{H} n_{h,t} \left\{\frac{E_{h,t}[p_{t+1}+y_{t+1}-Rp_t]}{a \sigma^2}\right\}=z_{s,t},
\end{equation}
where $H$ is the number of different investor types. In the simple case $H=1$ we obtain the equilibrium pricing equation and for the specific case of zero supply of outside risky shares, i.e. $z_{s,t}=0$ for all $t$, the market equilibrium then satisfies:
\begin{equation}\label{eq:g}
Rp_t=\sum_{h=1}^{H} n_{h,t}\{E_{h,t}[p_{t+1}+y_{t+1}]\}.
\end{equation}
In a completely rational market \autoref{eq:g} reduces to $Rp_t=E_t[p_{t+1}+y_{t+1}]$ and the price of the risky asset is completely determined by economic fundamentals and given by the discounted sum of its future dividend cash flow:
\begin{equation}\label{eq:h}
p_t^*=\sum_{k=1}^{\infty} \frac{E_t[y_{t+k}]}{(1+r)^k},
\end{equation}
where $p_t^*$ depends upon the stochastic dividend process $\{y_t\}$ and denotes the fundamental price which serves as a benchmark for asset valuation based on economic fundamentals under rational expectations. In the specific case where the process $\{y_t\}$ is independent and identically distributed, $E_t\{y_{t+1}\}=\bar{y}$ is a constant. The fundamental price, which all investors are able to derive, is then given by the simple formula:

\begin{equation}\label{eq:i}
p^*=\sum_{k=1}^{\infty} \frac{\bar{y}}{(1+r)^k}=\frac{\bar{y}}{r}.
\end{equation}
For the further analysis it is convenient to work not with the price levels, but with the deviation $x_t$ from the fundamental price $p_t^*$:
\begin{equation}\label{eq:j}
x_t=p_t-p_t^*.
\end{equation}

%***********************
\subsection{Heterogeneous beliefs}
\label{subsec:hetbel}

Now we introduce the heterogeneous beliefs about future prices. We follow the \cite{BrHo1998} approach and assume the beliefs of individual trader types in the form:
\begin{equation}\label{eq:k}
E_{h,t}(p_{t+1}+y_{t+1})=E_t(p_{t+1}^*+y_{t+1})+\mathnormal{f}_h(x_{t-1},\dots,x_{t-L}),~~~\textrm{for all}~h,t,
\end{equation}
where $p_{t+1}^*$ denotes the fundamental price (\autoref{eq:h}), $E_t(p_{t+1}^*+y_{t+1})$ denotes the conditional expectation of the fundamental price based on the information set
$\mathcal{F}_t=\{p_t,p_{t-1},\dots;y_t,y_{t-1},\dots\}$, $x_t=p_t-p_t^*$ is the deviation from the fundamental price (\autoref{eq:j}), $\mathnormal{f}_h$ is some deterministic function which can differ across trader types $h$ and represents a `$h$-type' model of the market, and $L$ denotes the number of lags.

It is now important to be very precise about the class of beliefs. From the expression in \autoref{eq:k} it follows that beliefs about future dividends flow:
\begin{equation}\label{eq:l}
E_{h,t}(y_{t+1})=E_t(y_{t+1}),~~~h=1,\dots H,
\end{equation}
are the same for all trader types and equal to the true conditional expectation. In the case where the dividend process $\{y_t\}$ is i.i.d., from \autoref{eq:i} we know that all trader types are able to derive the same fundamental price $p_t^*$.

On the other hand, traders' beliefs about future price abandon the idea of perfect rationality and move the model closer to the real world. The form of this class of beliefs:
\begin{equation}\label{eq:m}
E_{h,t}(p_{t+1})=E_t(p_{t+1}^*)+\mathnormal{f}_h(x_{t-1},\dots,x_{t-L}),~~~\textrm{for all}~h,t,
\end{equation}
allows prices to deviate from their fundamental value $p_t^*$, which is a crucial step in heterogeneous agent modelling. $\mathnormal{f}_h$ allows individual trader types to believe that the market price will differ from its fundamental value $p_t^*$.

An important consequence of the assumptions above is that heterogeneous market equilibrium from \autoref{eq:g} can be reformulated in the deviations form, which can be conveniently used in empirical and experimental testing. We thus use \autoref{eq:j}, \ref{eq:k} and the fact that $\sum_{h=1}^{H} n_{h,t}=1$ to obtain:
\begin{equation}\label{eq:n}
Rx_t=\sum_{h=1}^{H} n_{h,t} E_{h,t}[x_{t+1}]= \sum_{h=1}^{H} n_{h,t}\mathnormal{f}_h(x_{t-1},\dots,x_{t-L}) \equiv \sum_{h=1}^{H} n_{h,t}\mathnormal{f}_{h,t},
\end{equation}
where $n_{h,t}$ is the value related to the beginning of period $t$, before the equilibrium price deviation $x_t$ has been observed. The actual market clearing price $p_t$ might then be calculated simply using \autoref{eq:j} as $p_t=x_t+p_t^*$, expressed more precisely, combining \autoref{eq:i}, \autoref{eq:j}, and \autoref{eq:n} as:
\begin{equation}\label{eq:new}
p_t=x_t+p_t^*=\frac{\sum_{h=1}^{H} n_{h,t}\mathnormal{f}_{h,t}}{R}+\frac{\bar{y}}{r}.
\end{equation}

%***********************

\subsection{Selection of strategies}
\label{subsec:selstra}

Beliefs of individual trader types are updated evolutionary and thus create adaptive belief system, where the selection is controlled by endogenous market forces \citep{BrHo1997}. It is actually an expectation feedback system as variables depend partly on the present values and partly on the future expectations. Market fractions of trader types $n_{h,t}$ are then given by the discrete choice probability --- the multinomial logit model:

\begin{equation}\label{eq:q3}
n_{h,t}=\frac{\textrm{exp}(\beta U_{h,t-1})}{Z_t},
\end{equation}

\begin{equation}\label{eq:r}
Z_t\equiv \sum_{h=1}^{H} \textrm{exp} (\beta U_{h,t-1}),
\end{equation}

where the one-period-lagged timing of $U_{h,t-1}$ ensures that all information for the market fraction $n_{h,t}$ updating are available at the beginning of period $t$, $\beta$ is the intensity of choice parameter measuring how fast traders are willing to switch between different strategies. $Z_t$ is then normalization ensuring $\sum_{h=1}^{H} n_{h,t}=1$. 

%***********************
\subsection{Basic belief types}
\label{subsec:babety}

In the original paper by \citet{BrHo1998}, authors analyse the behaviour of the artificial market consisting of a few simple belief types (trader types or strategies). The aim of investigating the model with only two, three, or four belief types is to describe the role of each particular belief type in deviation from fundamental price and to investigate the complexity of the simple model dynamics with the help of the bifurcation theory.

All beliefs have the simple linear form:
\begin{equation}\label{eq:t}
\mathnormal{f}_{h,t}=g_h x_{t-1}+b_h,
\end{equation}
where $g_h$ denotes the trend and $b_h$ is the bias of trader type $h$. This form comes from the argument that only a very simple forecasting rules can have a real impact on equilibrium prices as complicated rules are unlikely to be learned and followed by sufficient number of traders. \citet{Hommes2006} also notices another important feature of \autoref{eq:t}, which is that $x_{t-1}$ is used to forecast $x_{t+1}$, because \autoref{eq:e} has not revealed equilibrium $p_t$ yet when $p_{t+1}$ forecast is estimated.

The first belief type are fundamentalists or rational `smart money' traders. They believe that the asset price is determined solely by economic fundamentals according to the EMH introduced in \citet{Fama1970} and computed as the present value of the discounted future dividends flow. Fundamentalists believe that prices always converge to their fundamental values. In the model, fundamentalists comprise the special case of \autoref{eq:t} where $g_h=b_h=\mathnormal{f}_{h,t}=0$. It is important to note that fundamentalists' demand also reflects market actions of other trader types. Fundamentalists have all past market prices and dividends in their information set $\mathcal{F}_{h,t}$, but they are not aware of the fractions $n_{h,t}$ of other trader types.

Chartists or technical analysts, sometimes called `noise traders' represent another belief type. They believe that asset price is not determined by economic fundamentals only, but it can be partially predicted using simple technical trading rules, extrapolation techniques or taking various patterns observed in the past prices into account. If $b_h=0$, trader $h$ is called a pure trend chaser if $0<g_h \leq R$ and a strong trend chaser if $g_h>R$. Additionally, if $-R \leq g_h<0$, the trader $h$ is called contrarian or strong contrarian if $g_h<-R$.

Next, if $g_h=0$ trader $h$ is considered to be purely upward biased if $b_h>0$ or purely downward biased if $b_h<0$.

\section{Empirical findings}

After the necessary introductions, we proceed to the analysis. In this section, we start with studying the behaviour of real world stock markets around a break point chosen to be a stock market crisis. In the next sections, we will try to answer the question if the empirical findings can be matched with the HAM. 

\subsection{Choice of the data}

The benchmark dataset we collect consists of all 30 constituents of the Dow Jones Industrial Average (DJIA) stock market index covering five particularly turbulent stock market periods. The sample we consider starts with Black Monday 1987, the largest one-day stock market drop in the history, and terminates with the Lehman Brothers Holdings bankruptcy in 2008, one of the milestones of the recent financial crisis of 2007--2010. 

While we analyse the dynamics of the model around the BPD where new behavioural elements are injected, data coming from the turbulent market periods will help us to verify or contradict our findings. Financial crises and stock market crashes can be widely considered as periods when investors' rationality is restrained and where behavioural patterns are likely to emerge, strengthen and often play the dominant role. For example \citet[pg. 72--76]{Malkiel2003} perceives periods surrounding Black Monday 1987 and Dot-com Bubble 2000 as typical examples of behavioural market crashes where not rational, \textit{``psychological considerations must have played the dominant role''}. \autoref{tab:events} gives a summary of particular events and related BPD.
\begin{table}[!htbp]\footnotesize
\caption{Financial Crises periods of 1987--2011}
\label{tab:events}
\centering
\begin{tabular}{p{4.1cm}lp{4.9cm}}
\hline
\noalign{\smallskip}
Event & \multicolumn{1}{c}{BPD} & \multicolumn{1}{c}{Description}
\\
\hline
\noalign{\smallskip}
Black Monday 1987 & October 19, 1987 (Mo) & The historical largest one-day DJIA drop: -22.61\%
\\
\hline
\noalign{\smallskip}
Ruble Devaluation 1998 & August 17, 1998 (Mo) & Russian government announced the ruble devaluation and ruble-denominated debts restructuring; important point in the Russian financial crisis
\\
\hline
\noalign{\smallskip}
Dot-com Bubble Burst & October 10, 2000 (Fri) & NASDAQ reached its historical intra-day peak of 5132.52 USD and closed at 5048.62; on Monday 13 it opened, however, at 4879.03, recording the 3\% decline
\\
\hline
\noalign{\smallskip}
WTC 9/11 Attack in NY & September 11, 2001 (Tue) & NYSE  consequently closed and reopened again after September 17 (Mo)
\\
\hline
\noalign{\smallskip}
Lehman Brothers Holdings Bankruptcy & September 15, 2008 (Mo) & `Chapter 11' bankruptcy protection publicly announced; one of the milestones of the Financial crisis of 2007-2010\\
\hline
\end{tabular}
\begin{center}
\end{center}
\end{table}
%\begin{figure}[!htbp]
%\caption{Financial Crises --- DJIA index prices (USD) and returns (\%). The time frame of all x-axes is +/- 20 working days from the depicted BPD.}
%\label{fig:crises}
%\vspace{1em}
%\centering
%  \subfloat[Black Monday 1987 --- Prices]{
%		\includegraphics[width=0.38\textwidth]{Figures/87price}}
%  \subfloat[Black Monday 1987 --- Returns]{
%		\includegraphics[width=0.38\textwidth]{Figures/87return}}
%\\
%  \subfloat[Ruble Devaluation 1998 --- Prices]{
%		\includegraphics[width=0.38\textwidth]{Figures/98price}}
%  \subfloat[Ruble Devaluation 1998 --- Returns]{
%		\includegraphics[width=0.38\textwidth]{Figures/98return}}
%\\
%  \subfloat[Dot-com Bubble 2000 --- Prices]{
%		\includegraphics[width=0.38\textwidth]{Figures/2000price}}
%  \subfloat[Dot-com Bubble 2000 --- Returns]{
%		\includegraphics[width=0.38\textwidth]{Figures/2000return}}
%\\
%  \subfloat[WTC 9/11 Attack 2001 --- Prices]{
%		\includegraphics[width=0.38\textwidth]{Figures/2001price}}
%  \subfloat[WTC 9/11 Attack 2001 --- Returns]{
%		\includegraphics[width=0.38\textwidth]{Figures/2001return}}
%\\
%  \subfloat[Lehman Brothers 2008 --- Prices]{
%		\includegraphics[width=0.38\textwidth]{Figures/2008price}}
%  \subfloat[Lehman Brothers 2008 --- Returns]{
%		\includegraphics[width=0.38\textwidth]{Figures/2008return}}
%\begin{center}
%\end{center}
%\end{figure}
We try to minimize the arbitrariness of event choice by considering two reasonable criteria which chosen events ought to satisfy. First, the BPD should be well determinable. A good example is e.g. September 11, 2001. Second, when looking back, the event should have had a direct and substantial impact on the U.S. stock market. Therefore, e.g. Black Wednesday in the UK in 1992 is not considered.

Several financial crisis periods such as well known Scandinavian banking crises in 1990s, already mentioned Black Wednesday in 1992, Asian financial crisis around 1997, and others do not satisfy some of these criteria well and therefore we omit them from our analysis. Moreover, to obtain a mutually consistent data, we have purposely decided to use the DJIA index, one of the oldest, closely watched, and most stable indices in the world. The need for certain consistency also restricts how deep into history we can immerse. With regard to historical changes among DJIA components, March 12, 1987 has been assessed as the historical border line as since then half of the index components have been gradually substituted. Detailed description of the DJIA components is given later in the text. Additionally, if there are more potential BPD during a particular crisis period, we opt for the first one as we suppose this one is likely to have the greatest psychological impact and even if not, it fits best into the model we consider in this work.

\subsection{Data description}
\label{sec:dataset}

The sample consists of the differences of the daily closing prices for all DJIA stock market index constituents covering five particularly turbulent stock market periods (\autoref{tab:events}). The price differences have been chosen instead of usually analysed stock prices or returns as they empirically resemble the outcomes of the model under the study --- after repeated simulations the average statistics are mutually comparable, especially in the short-run.

We choose one month (20 working days) before and one month (20 working days) after the specific BPD. We have decided for this rather short period for two reasons. First, the dynamics of the model stabilizes quite fast. Second, from the psychological aspect, we expect to observe the most interesting changes in behavioural patterns especially very close to the BPD. The full sample consists of 5519 observations. 

Some of the historical data are, nonetheless, unavailable for particular stocks. The reason is that either the company ceased to exist (a bankruptcy, a merge, or an acquisition) or a particular stock trading was suspended for some time. In both cases the historical data of a particular stock are unavailable.\footnote{If a company changed its name, it usually keeps its ticker symbol and the historical data remain available.} There is also one specific case of the American International Group (AIG) in the dataset. AIG stock was removed from the index and replaced by the Kraft Food (KFT) on September 22, 2008 --- i.e. during the `Lehman Brothers 2008' period --- after a series of enormous drops around September 15, 2008. To keep sample consistency we omit this stock as an outlier.

DJIA index consists of 30 stock and the index base has changed 48 times during its history starting in 1896. Therefore the structure of the partial samples varies across time and due to unavailability of some data, the partial samples are not all of the same magnitude. However, for the purpose of our analysis this is not a problem. What we only need is a consistent set of as many observations as possible to extract patterns of data dynamics around the BPD. \autoref{tab:stocks} in Appendix outlines changes in the structure of the index during the period we consider and shows precisely which data are used.

\begin{figure}[h]%[!htbp]
\caption{DJIA price differences (USD) distributions. Full lines depict empirical PDFs of data before and after (gray filling) the BPD. Discontinuous lines depict fits of N($\mu$,$\sigma^2$) before (dotted) and after (dashed) the BPD. For comparison, the PDFs of the normal distribution based on each particular sample mean and standard deviation are depicted.}
\label{fig:beforeafter1}
\vspace{1em}
\centering
  \subfloat[Black Monday 1987]{
   \includegraphics[width=0.5\textwidth]{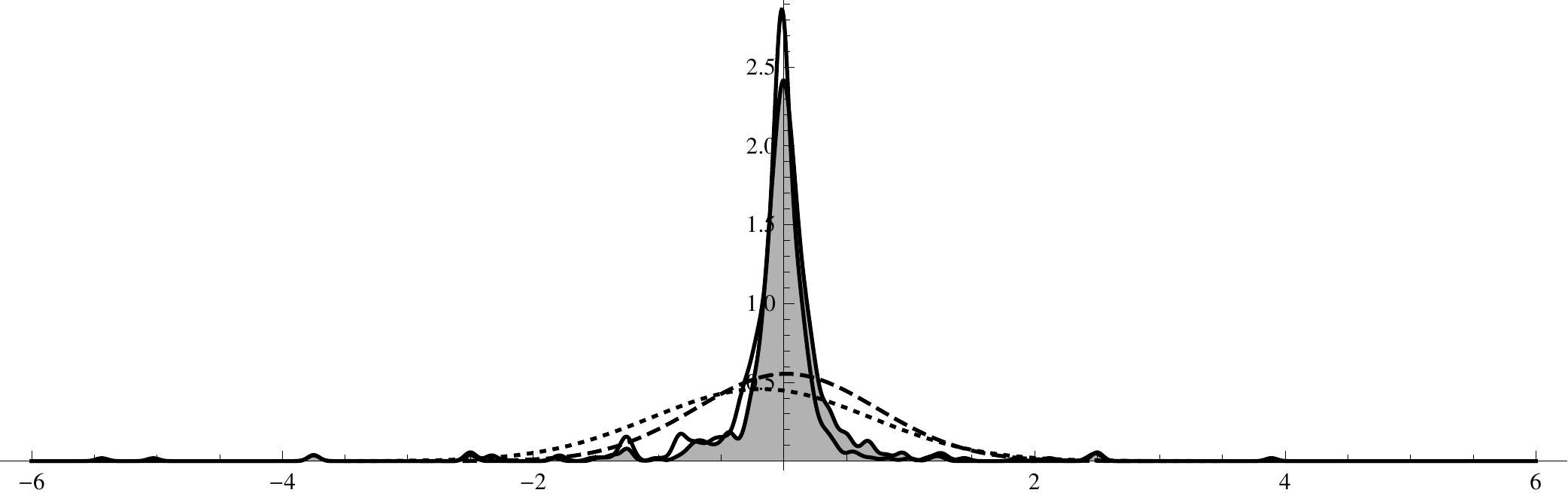}}
  \subfloat[Ruble Devaluation 1998]{
   \includegraphics[width=0.5\textwidth]{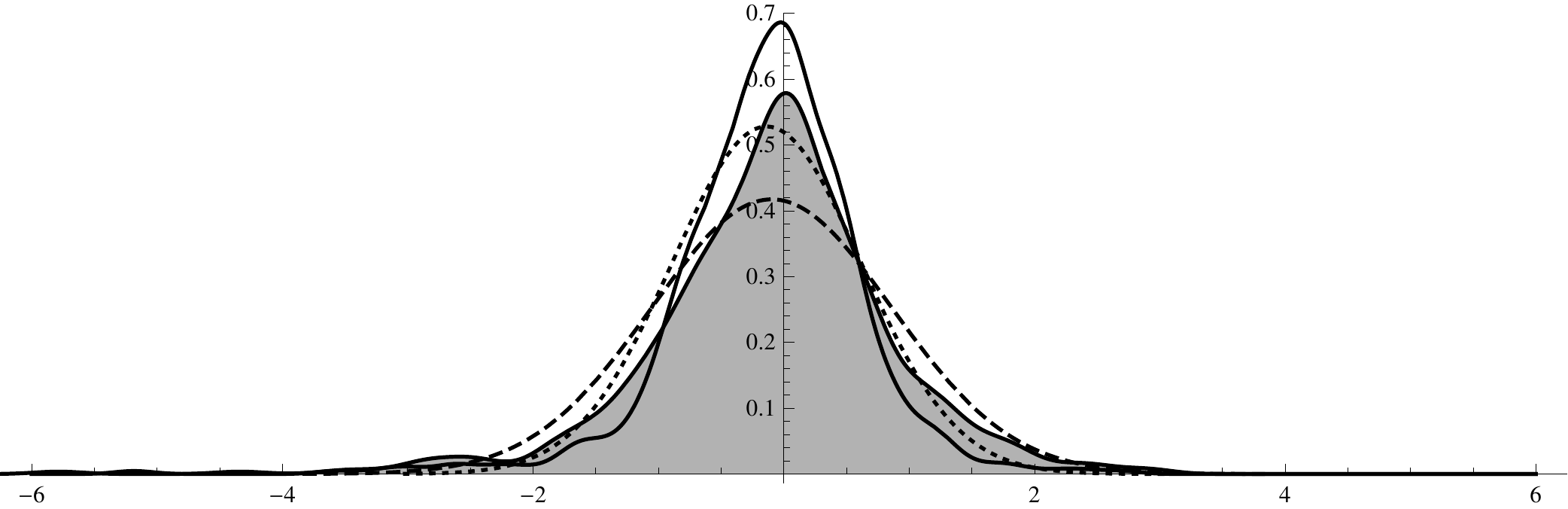}}
\\
  \subfloat[Dot-com Bubble 2000]{
   \includegraphics[width=0.5\textwidth]{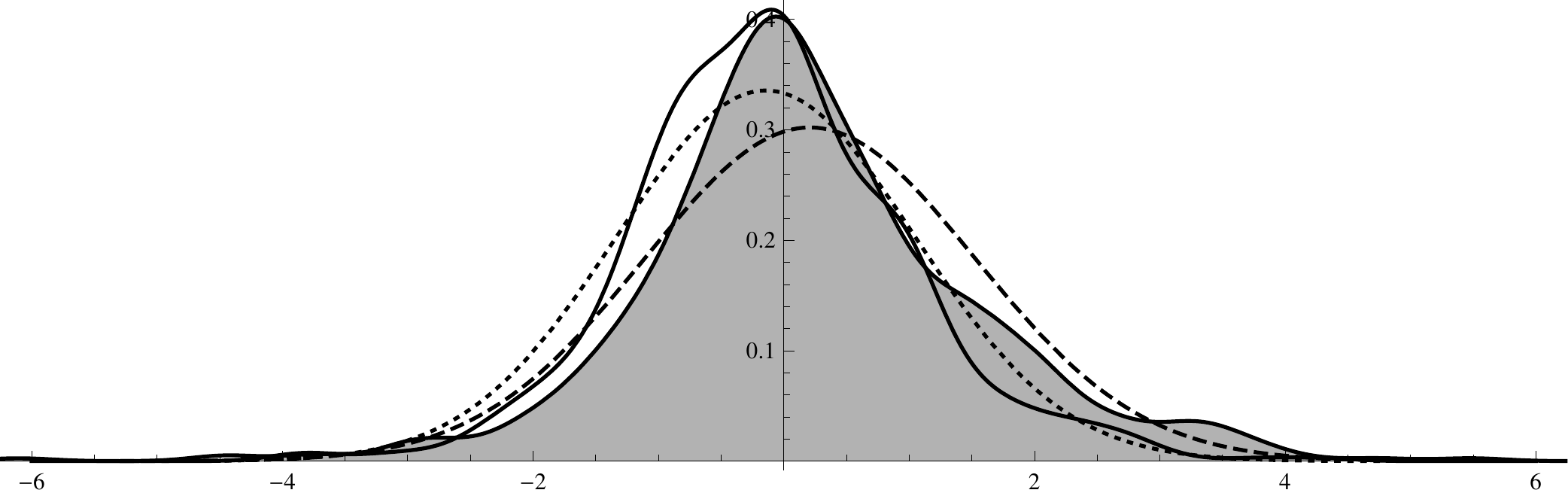}}
  \subfloat[WTC 9/11 Attack 2001]{
   \includegraphics[width=0.5\textwidth]{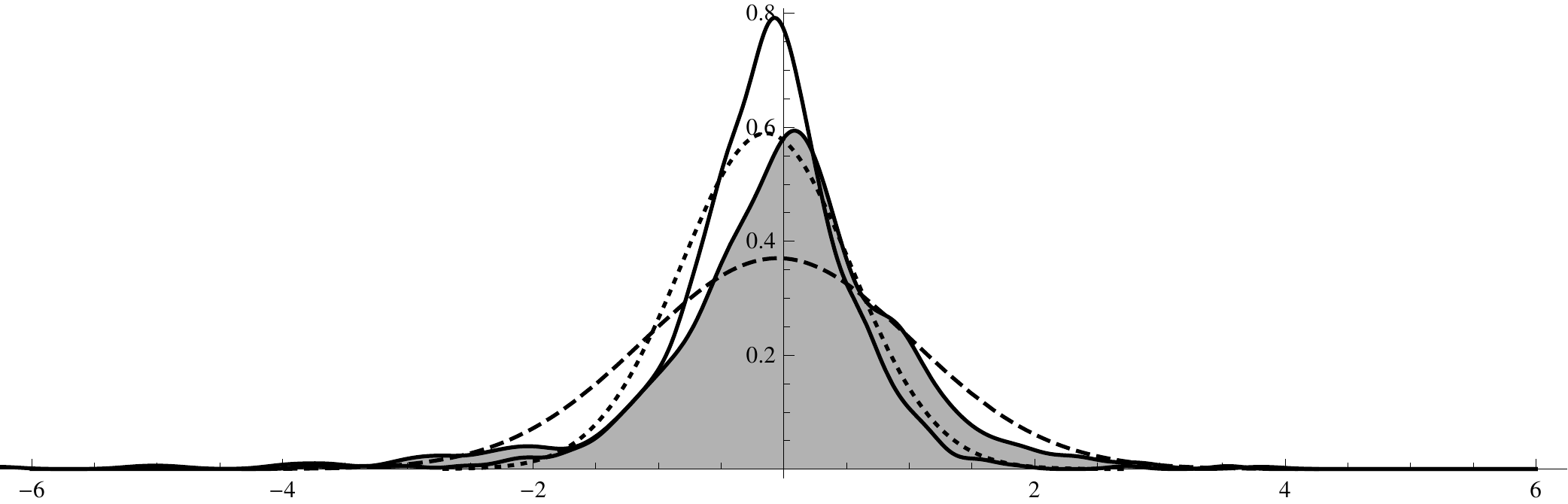}}
\\
  \subfloat[Lehman Brothers 2008]{
   \includegraphics[width=0.5\textwidth]{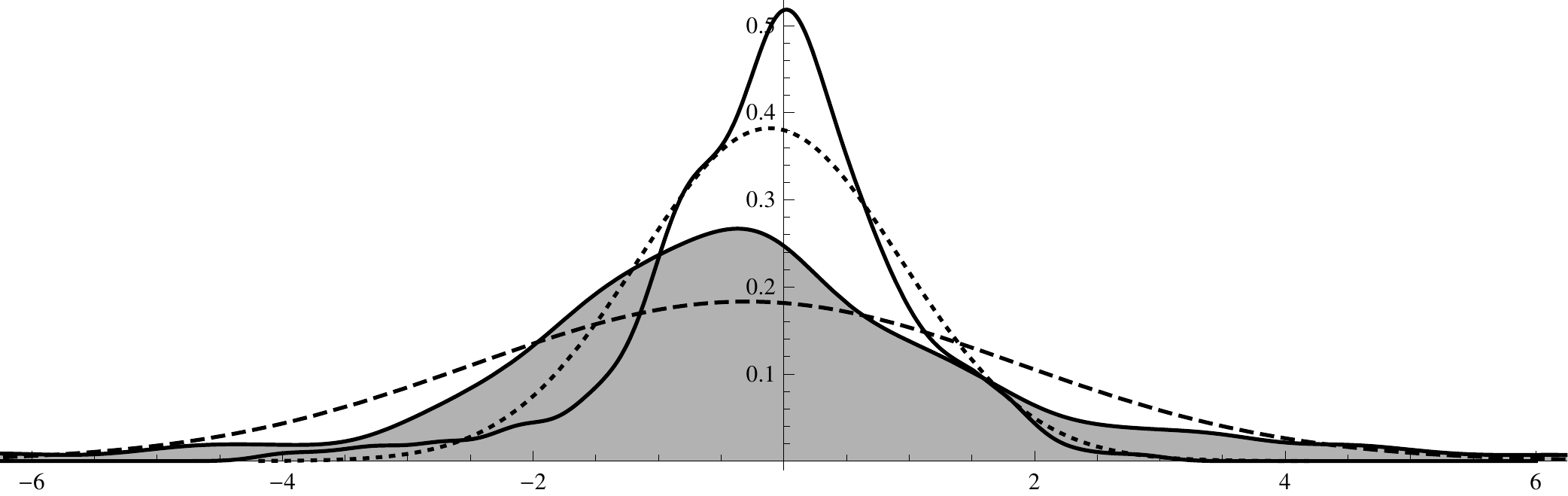}}
\begin{center}
\bigskip
\end{center}
\end{figure}

\begin{figure}[!htbp]
\caption{Tails of the aggregated DJIA constituents' price differences (in USD) empirical distributions. Left tail is marked with ``$+$", right tail with ``$o$", while tails before the BPD is depicted in black and tails after the BPD is depicted in gray.}
\label{fig:beforeafter}
\vspace{1em}
\centering
  \subfloat[Black Monday 1987]{
   \includegraphics[width=0.4\textwidth]{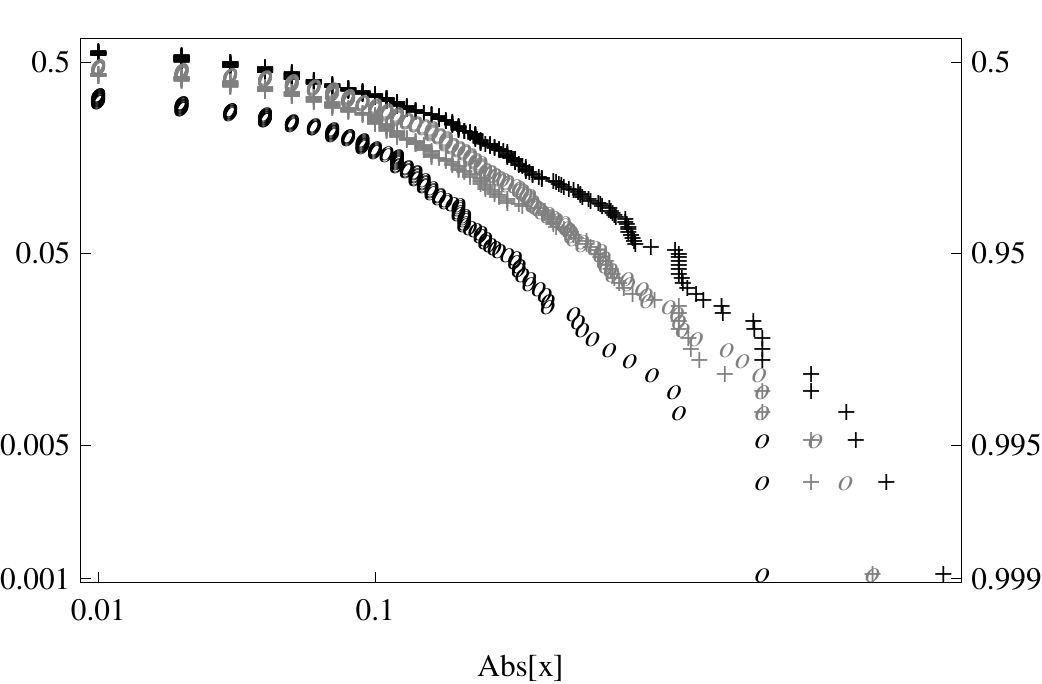}}
  \subfloat[Ruble Devaluation 1998]{
   \includegraphics[width=0.4\textwidth]{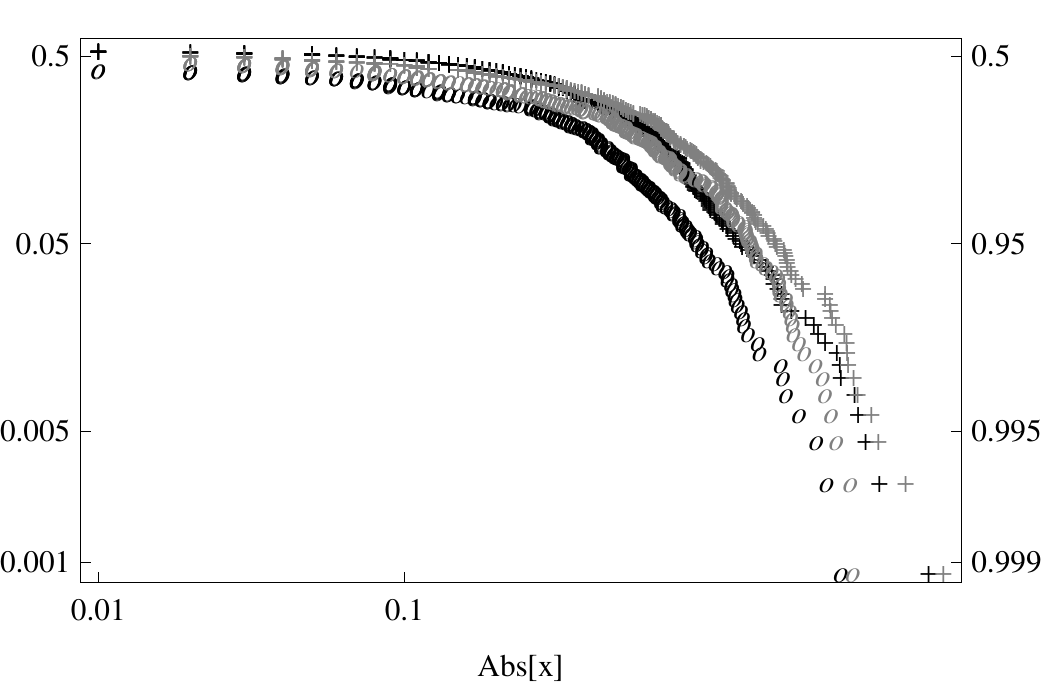}}
\\
  \subfloat[Dot-com Bubble 2000]{
   \includegraphics[width=0.4\textwidth]{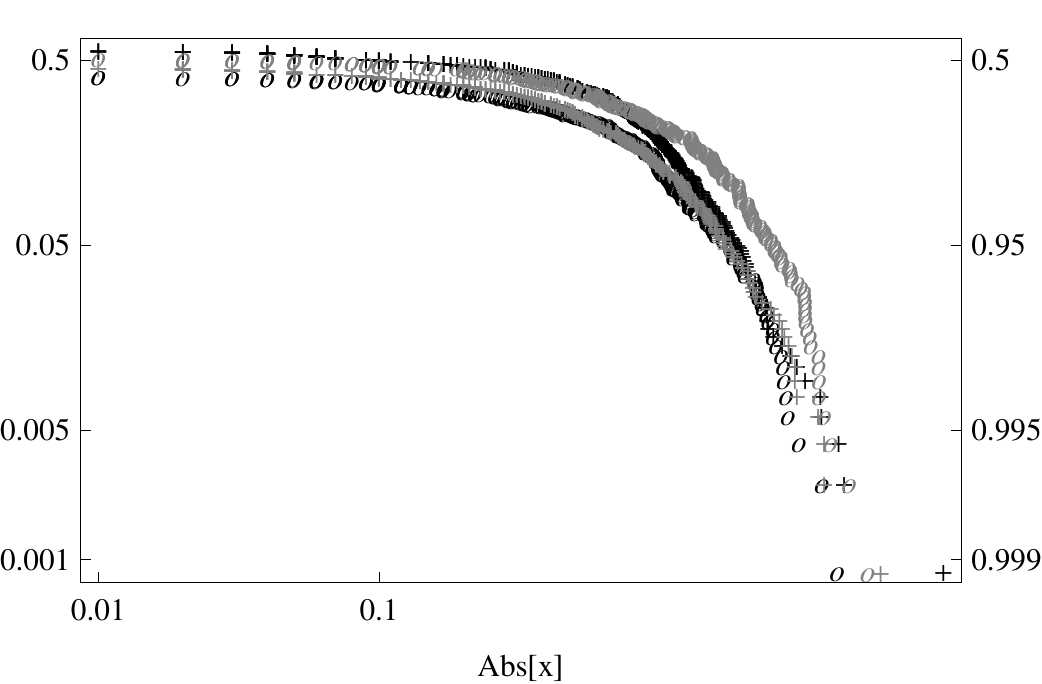}}
  \subfloat[WTC 9/11 Attack 2001]{
   \includegraphics[width=0.4\textwidth]{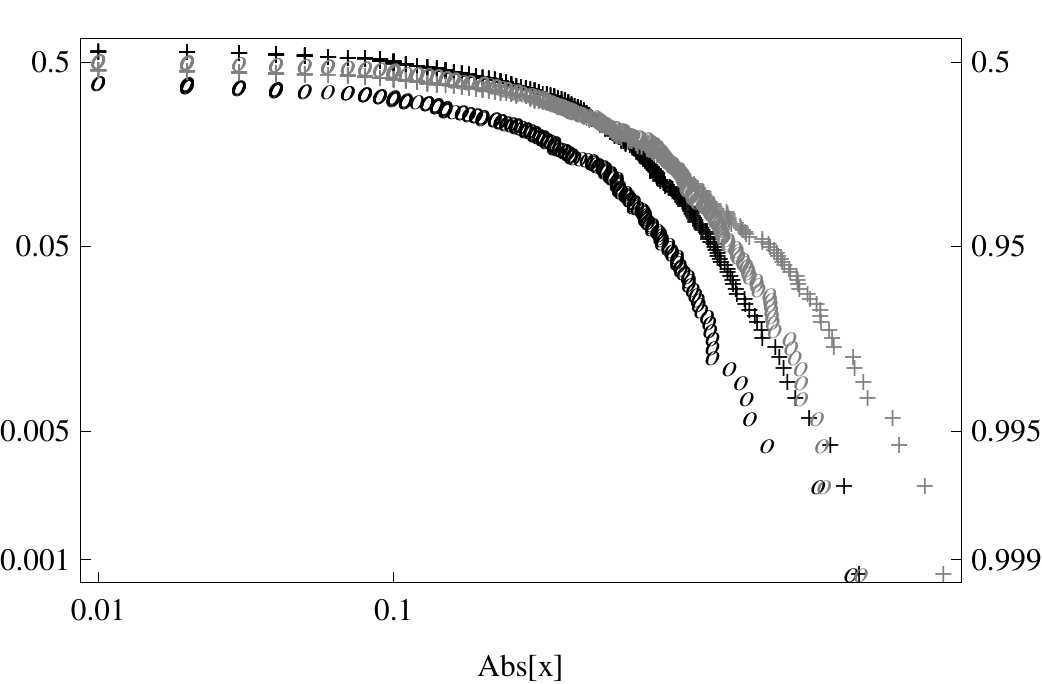}}
\\
  \subfloat[Lehman Brothers 2008]{
   \includegraphics[width=0.4\textwidth]{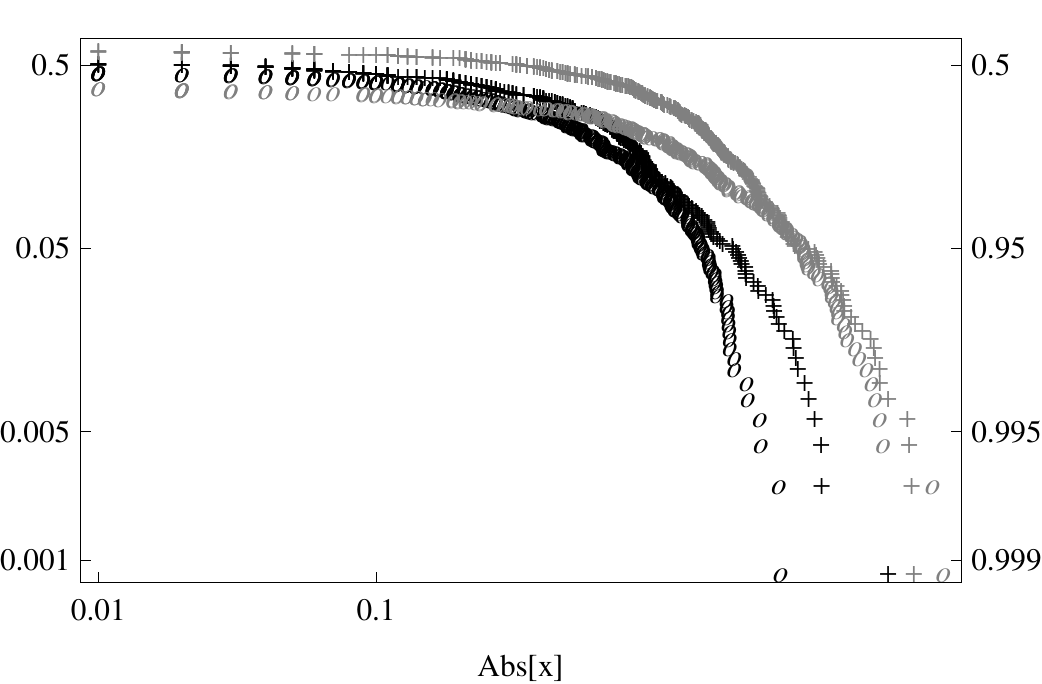}}
\begin{center}
\bigskip
\end{center}
\end{figure}

\autoref{fig:beforeafter1} depicts probability density functions of the empirical distributions of the price differences for each partial sample before and after the BPD. Excess kurtosis and heavy tails are distinct at first glance. \autoref{fig:beforeafter} shows tails of the empirical distributions of the price differences for each partial sample before and after the BPD. \autoref{tab:desstats} then offers a summary of important descriptive statistics keeping the same logic as \autoref{fig:beforeafter1} and \autoref{fig:beforeafter}.

\begin{sidewaystable}[!htbp]
\caption{Partial samples descriptive statistics. Note that B/A denotes `Before/After' the Break Point Date. $\Delta$ is the change in magnitude B/A. Skew. denotes skewness, Kurt. denotes kurtosis, JB denotes the Jarque--Bera Test for normality of distribution. Asterisks denote statistical significances of equal means xor variances of B/A based on the mean difference test and the variance ratio test: $^{***}$ denote hypothesis rejected at 1\% significance level, $^{**}$ at 5\% level, $^{*}$ at 10\% level.}
\label{tab:desstats}
\centering
\begin{tabular}{lcrlrlrlrlrrr} %pìkné zarovnání
\hline
\noalign{\smallskip}
~~Event & B/A & Mean & $\Delta$ & Variance & $\Delta$ & Skew. & $\Delta$ & Kurt. & $\Delta$ & Min. & Max. & JB \\
   &     & (USD) &  & (USD) & (\%) &  &  &  & \multicolumn{1}{c}{(\%)} & (USD) & (USD) & \multicolumn{1}{c}{(p-val.)}\\
\hline
\noalign{\smallskip}
~~Black Monday        & B & -0.181 &               & 0.762 &                 & -6.689 &                & 72.130 &                 & -11.25 & 2.50 & 0.000 \\
                      & A & 0.022   & \textuparrow$^{***}$   & 0.484 & \textdownarrow$^{***}$~31.9 & 0.484   & \textuparrow   & 37.126 & \textdownarrow~48.5 & -6.25  & 6.25 & 0.000\\

~~Ruble Devaluation    & B & -0.136 &               & 0.571 &                       & -1.017 &                & 8.716  &                 & -5.17  & 2.67 & 0.000 \\
                      & A & -0.096 & \textuparrow$^{}$   & 0.914 & \textuparrow$^{***}$~60.0   & -0.721 & \textuparrow   & 6.566  & \textdownarrow~24.7 & -5.77  & 2.92 & 0.000\\

~~Dot-com Bubble      & B & -0.143 &                & 1.413 &                       & -0.967 &                & 12.406 &                 & -10.24 & 4.29 & 0.000\\
                     & A & 0.208   & \textuparrow$^{***}$   & 1.744 & \textuparrow$^{***}$~23.4   & 0.178   & \textuparrow   & 4.710  & \textdownarrow~62.0 & -6.12  & 5.50 & 0.000\\

~~WTC 9/11 Attack     & B & -0.142 &                & 0.458 &                       & -0.376  &                & 7.554  &                 & -3.76  & 3.54 & 0.000 \\
                     & A & -0.046 & \textuparrow$^{**}$   & 1.162 & \textuparrow$^{***}$~153.6  & -1.471 & \textdownarrow & 10.328 & \textuparrow~36.7   & -7.25  & 3.82 & 0.000\\
 
~~Lehman Brothers     & B & -0.112 &                & 1.090 &                       & -0.986 &                & 7.234  &                 & - 6 .97 & 2.86 & 0.000\\
                     & A & -0.295 & \textdownarrow$^{**}$ & 4.744 & \textuparrow$^{***}$~335.3  & 0.273   & \textuparrow   & 6.603  & \textdownarrow~8.7 & -8.63   & 11.02 & 0.000 \\
\hline
\noalign{\smallskip}
~~Summary &&& 4/5 \textuparrow && 4/5 \textuparrow$^{***}$ && 4/5 \textuparrow && 4/5 \textdownarrow &&&
\\
\bottomrule
    \end{tabular}
\begin{center}
\end{center}
\end{sidewaystable}

To check the robustness of our findings, we have also developed three analogous datasets to compare the results of the primal sample. For the first control dataset we consider the same structure of stocks but the observed period is twice as long., i.e. 40 working days before and after the BPD. Therefore the sample consists of double amount of data. The second and third control sets comprise only those stock that remained in the index for the entire observed period (AA, AXP, BA, DD, GE, GM, IBM, KO, MCD, MMM, MRK, PG, T, UTX, XOM) and we again analyse the 20 and 40 working days before and after the BPD. Given the very same set of stocks considered in each period all partial subsamples then have the same magnitude, i.e. 300 and 600 observations, respectively. The comparison does not reveal, however, any significant difference from the original dataset which makes us conclude that our test sample is chosen robustly.

\subsection{What can we infer from the data?}
\label{sec:infer}

At first glance, our data does not come from Gaussian normal distribution, which is confirmed by the Jarque--Bera Test for normality of distribution (see \autoref{tab:desstats}). Typical stylized facts of financial returns such as excess kurtosis or heavy tails are also fulfilled. In \autoref{tab:desstats} one can clearly see shifts of mean, variance, skewness and kurtosis between the `before' and `after' periods. The first three of these four descriptive statistics increase in four of five analysed periods, while kurtosis decreases in the same ratio of cases. These findings are not only interesting from the statistical point of view, economic interpretation might be interesting to study as well.

Let us start with mean which rather increases. One of the possible explanations might be that after a sudden market crash the resulting short-run tendency is to compensate the huge drop in prices by several increases. Speculators might play a substantial role in this situation. These increases then statistically exceed the huge drop at the beginning which leads to the overall picture of a higher mean. Another rationale might be that the market crash is a climax when the negative trend of market prices culminates. After that, therefore, there is no more space for drops and the market naturally increases.

Increasing variance measures the increasing market risk and uncertainty resulting in its unpredictability. This is one of the accompaniments of all turbulent market periods.

The most challenging issue for an economic interpretation is the increasing skewness. This means that the mass of the distribution shifts from right to left, the right tail becomes longer and high values become more scarce. All of these features are likely to be related to general crisis tendencies but their interpretation with regard to the increasing mean does not seem unambiguous.

Finally, for the decreasing trend of kurtosis several straightforward explanations may apply. In the crisis period after a market crash extreme observations (both negative and positive) become more likely and, in the contrary, observations close to mean are less probable. Tails of related distribution thus become heavier as shown by \autoref{fig:beforeafter1} and \autoref{fig:beforeafter} and kurtosis decreases.

The most interesting part now will be to compare these considerable empirical results with the outcomes of the model simulations.

%********

\section{A setup for simulations}
\label{sec:jsett}

To be able to mutually compare all different model setups, we define a joint setup which will be used for all simulations. In the simulations, we follow our previous work \citep{BaVaVo2009,VaBaVo2009,VaBaVo2010} basing our model in the \citet{BrHo1998} setting. Adaptive belief system is compactly described \citep{Hommes2006} by the three mutually dependent equations:
\begin{eqnarray}
\label{eq:setting1}
R x_t     &=& \sum_{h=1}^{H} n_{h,t}\mathnormal{f}_{h,t}+ \epsilon_t \equiv \sum_{h=1}^{H} n_{h,t}(g_h x_{t-1}+b_h)+ \epsilon_t,\\
\label{eq:setting3}n_{h,t}   &=& \frac{\textrm{exp}(\beta U_{h,t-1})}{\sum_{h=1}^{H} \textrm{exp} (\beta U_{h,t-1})}, \\
U_{h,t-1} &=& (x_{t-1}-Rx_{t-2})\frac{\mathnormal{f}_{h,t-2}-Rx_{t-2}}{a\sigma^2}\nonumber \\
          &\equiv& (x_{t-1}-Rx_{t-2})\frac{g_h x_{t-3}+b_h-Rx_{t-2}}{a\sigma^2},
\label{eq:setting4}
\end{eqnarray}
where $\epsilon_t$ denotes the noise term representing the market uncertainty and unpredictable occasions. The Walrasian price formation mechanism is inherently comprised within the first \autoref{eq:setting1} of this system as described in detail in \autoref{sec:model} and summarized in \autoref{eq:n} and \autoref{eq:new}.

 The inevitable feature of all heterogeneous agent models are too many degrees of freedom together with a large number of parameters which can be modified and studied. Therefore we need to fix several variables to be able to analyse particular changes of the model ceteris paribus. As in \citet{BaVaVo2009} and \citet{VaBaVo2009}, we set the constant gross interest rate $R=1+r=1.1$; the linear term $1/{a\sigma^2}$ consisting of the risk aversion coefficient $a>0$ and the constant conditional variance of excess returns $\sigma^2$ is fixed to $1$. In addition to that, we use relatively small number of traders, $H=5$ and neither memory nor learning process are implemented to keep the impact of the behavioural modifications as clear as possible.

To examine the impact of suggested changes on the model outcomes, we rely on Monte Carlo methods. To obtain statistically valid and reasonably robust sample, we rely on the number of 100 runs. The trend parameter $g_h$ is drawn from the normal distribution $N(0,0.16)$,
the bias parameter $b_h$ is drawn from the normal distribution $N(0,0.09)$. When $g_1=b_1=0$, fundamentalists are present at the market. We allow fundamentalists to appear in the market either randomly or by control.

Finally, the magnitude of noise has to be considered carefully as it represents the notion of market uncertainty so it is inevitable part of the model, but it should not overshadow the effect of analysed modifications. We examine the effect of various noise settings, namely $\epsilon_t \in U(-0.02,0.02)$, $\epsilon_t \in U(-0.05,0.05)$ and $\epsilon_t \in U(-0.1,0.1)$ and conclude that although different noise variance causes some minor changes in model outcomes, all models across different noises embody major similarities. Thus the noise term $\epsilon_t$ is drawn from the uniform distr. $U(-0.05,0.05)$ which seem reasonable\footnote{Results for different noise settings are available upon request from authors.} to us.

%***********************
%\subsection{Algorithm description}
%\label{sec:al}

In our simulations, we change the intensity of choice $\beta$ and the intensity of the behavioural element. The literature estimating $\beta$ using a real marked data is extremely scarce as it is difficult to estimate the intensity of choice due to the non-linear nature of the model. One recent example is \citet[pg. 2281]{FrLeZw2010} who find that the intensity of choice \textit{``is positive and of considerable magnitude throughout the sample''} of daily closing DAX prices covering the entire year 2000. Hence $\beta$ still remains a rather theoretical concept. However, larger $\beta$ implies higher willingness of traders to switch between strategies based on their profitability --- the best strategies at each specific period are chosen by more agents --- and reversals in the price development are thus more likely. To comprise the large variety of possible values we use\footnote{In the choice of $\beta$, we follow our previous research. In \citet{BaVaVo2009} and \citet{VaBaVo2009}, we use $\beta=300$, in \citet{VaBaVo2010}, we use $\beta=500$ for example. Results for different $\beta \in \langle 0,\ldots  1000\rangle$ are available upon request from authors.}  the range $\beta \in \langle 5,\ldots  500\rangle$ with single steps of $55$, i.e. $\beta=\{5, 60, 115, 170, 225, 280, 335, 390, 445, 500\}$.  On the other hand, for each behavioural element the range varies and we describe its logic in further sections. The length of the generated time series corresponds to 250 days, i.e. $t=\{0,\ldots 250\}$. First 10\% of observations are discarded as initial period.

The way we implement the behavioural elements into the framework is that we change the dynamics of the model in the middle of generated time series, i.e. from the 126th iteration via injecting a new behavioural element. Then, we study the dynamics of the system closely before and after this change. The choice of the empirical benchmark sample introduced in the previous text reflects exactly this procedure. The last 10\% of observations are discarded as well to obtain samples of the same magnitude.

In studying the effect of behavioural elements on the model outcomes, we focus on four sub-series. First, we split the generated series into 2 halves, before and after the 126 iteration when behavioural break is put into the model. Second, we focus on last 20 observations before the behavioural break and first 20 observations after the behavioural break. By cumulating these four groups of series from different iterations, we are finally left with (i) the complete `before' sample of 10000 observations; (ii) the 20 day `before' sample of 2000 observations; (iii) the 20 day `after' sample of 2000 observations; (iv) the complete `after' sample of 10000 observations.

%***********************
 
\subsection{Modelling of behavioural patterns: herding}
\label{sec:behel}

As a certain notion of herding is naturally included\footnote{See e.g. \citet{ChHe2002}, \citet{ChGaLePa2003}, \citet{DeGr2006}, or \citet{Hommes2006}.} in the evolutionary adaptive system of strategy switching, we present different original modelling approach to herding. The examination of herding patterns in HAM is always based on short-run profitabilities of individual strategies and herding is detected via the evolution of market fractions $n_{h,t}$. This concept of herding is hence more or less (boundedly) rational. Therefore we introduce a concept of rather irrational, `blind' herding, which is based on public information and aims to imitate traders' behaviour during large stocks sell-offs after a market crash. 

In this approach one of trading strategies ($h=5$) does not behave in the traditional way but copies the behaviour of the most successful traders of the previous day. At time $t$ the strategy primarily evaluates its own performance measure, then compares the performance measures of all other strategies and for the next period $t+1$ it adjusts its beliefs about the trend $g_5$ and bias $b_5$ parameters, so that they mimic the last period's most profitable strategy. The mimicking effect is thus one period lagged and this delay represents the reaction of less informed market participant who just follow the crowd. 

Thus instead of 5 strategies present at the market, we can observe only 4 of them while the last, fifth mimics the most profitable one. Including the fifth mimicking strategy can lead to substantially different results in favor of the strategy which is currently being imitated, especially when the intensity of choice $\beta$ is small.

\subsection{Modelling of behavioural patterns: overconfidence}

Behavioural overconfidence can be modelled as a routine tendency to overestimate the accuracy of own judgments. Trying to incorporate this into the model framework, we are left with no other choice than work with the trend $g_h$ and bias $b_h$ parameters. However, this makes perfect sense and we model overconfidence as an overestimation of generated values. Roughly speaking, an overconfident trend chasing trader behaves even more surely and follows the observed trend strongly than in a normal (randomly generated) situation. He also expects the price to rise or drop even more than according to his (randomly generated) premises. The range of the `overconfidence element' is $\langle 0.05,\ldots  0.5\rangle$ and one can imagine this as the representation of the excess assurance in percentage terms --- from 5 to 50\%. Three options are examined: overconfidence affecting the trend parameter $g_h$ only, overconfidence affecting the bias parameter $b_h$ only, and overconfidence affecting both parameters. As overconfidence is positive `from definition', negative values are not considered.

\subsection{Modelling of behavioural patterns: market sentiment}

We model the market sentiment as shifts of the mean values of probability distributions from which the trend parameter $g_h$ and the bias parameter $b_h$ are generated. Both impacts of the `positive' and the `negative' sentiment are examined. \citet{VaBaVo2009} model the market sentiment as jumps of the trend parameter $g_h$ between realizations from the normal distributions $N(0.04,0.16)$ and $N(-0.04,0.16)$. We generally consider four options: (i) sentiment affecting the trend parameter $g_h$ only (the same case as we consider in \citealp{VaBaVo2009}); (ii) sentiment affecting the bias parameter $b_h$ only; (iii) sentiment affecting both parameters; (iv) and finally so called `mixed' case where the positive sentiment affecting bias $b_h$ is combined with the negative sign of the trend parameter $g_h$ and vice versa. The interpretation of both effects is, however, considerably different. 

If we decrease the mean of the trend parameter $g_h$, the contrarian strategies are more likely to be generated. Nonetheless, this does not tell much about the type of sentiment we have set --- whether the sentiment is intended to be positive or negative --- we primarily adjust the response of agents to actual price development. On the other hand, manipulating with the mean of the bias parameter $b_h$ we directly set the trend --- decreasing the mean we model negative market sentiment (price rather drops tomorrow) and vice versa. Next, as we aim to model the behaviour in extremely turbulent times, we assume higher shifts than only one tenth of the standard deviation as in \citet{VaBaVo2009}. The range for the positive sentiment element is $\langle 0.04,\ldots,0.4\rangle$ for trend and $\langle 0.03,\ldots,0.3\rangle$ for bias, i.e. the minimum is one tenth of the standard deviation and the maximum equals one standard deviation of related distributions. For degree of negative sentiment, the opposite values $\langle -0.4,\ldots,-0.04\rangle$ for trend and $\langle-0.3,\ldots,-0.03\rangle$ for bias are used. At first sight, this approach might seem a little similar to the overconfidence modelling, but the opposite is true. Although both appertain to the trend and bias parameters, in the overconfidence case we only symmetrically strengthen traders' responses to the current market development, while in the market sentiment case we asymmetrically deflect the market behaviour and traders' beliefs.

\subsection{Illustration of generated series}

\begin{figure}[h]%[!htbp]
\caption{Illustration of generated data for different behavioural breaks in the model. The upper part shows the data simulated from original model, bottom part depicts counterpart of the simulation where (d) herding, (e) overconfidence, and (f) market sentiment is injected.}
\label{fig:typical}
\vspace{1em}
\centering
  \subfloat[Original series 1]{
   \includegraphics[width=0.33\textwidth]{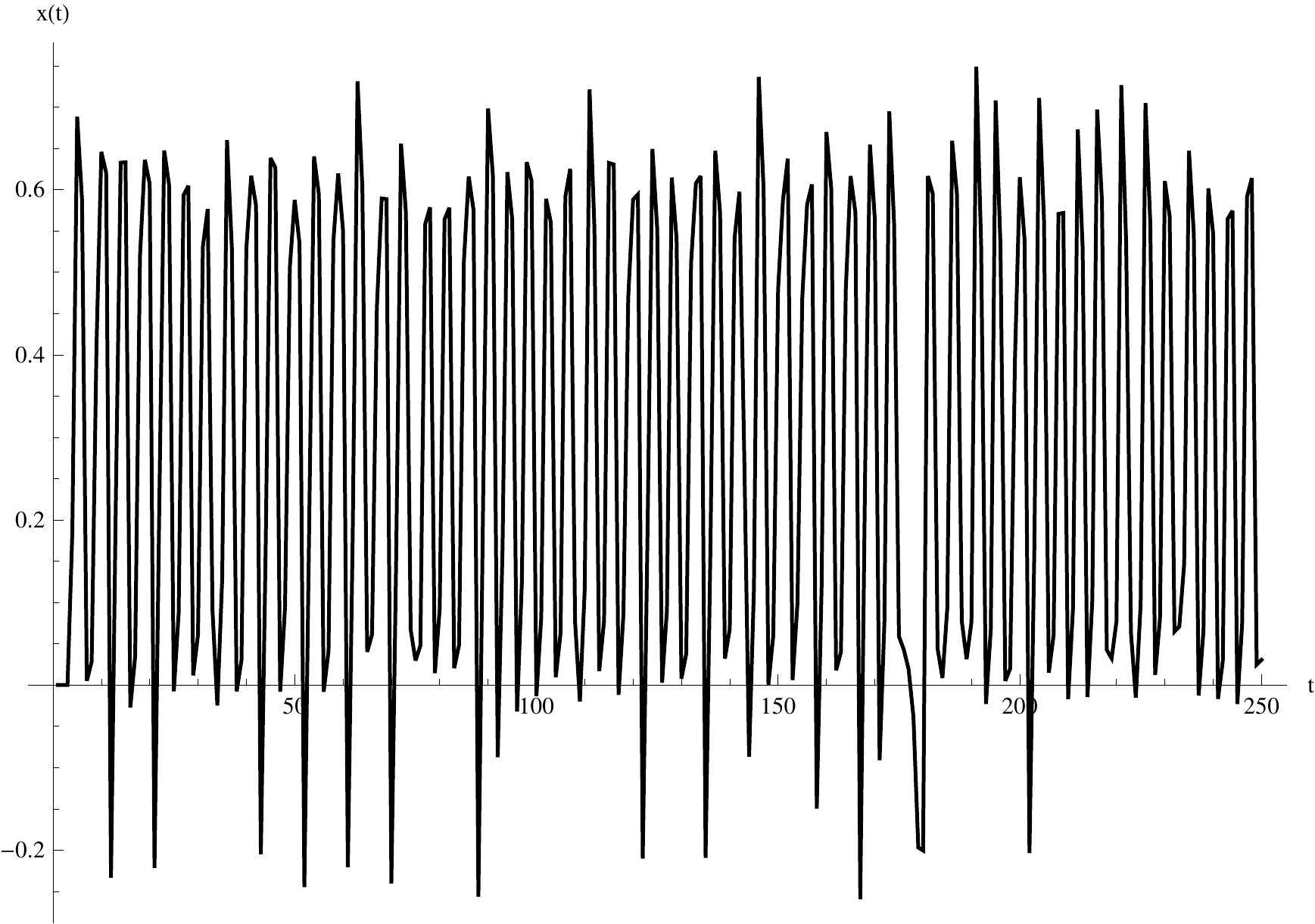}}
   \subfloat[Original series 2]{
   \includegraphics[width=0.33\textwidth]{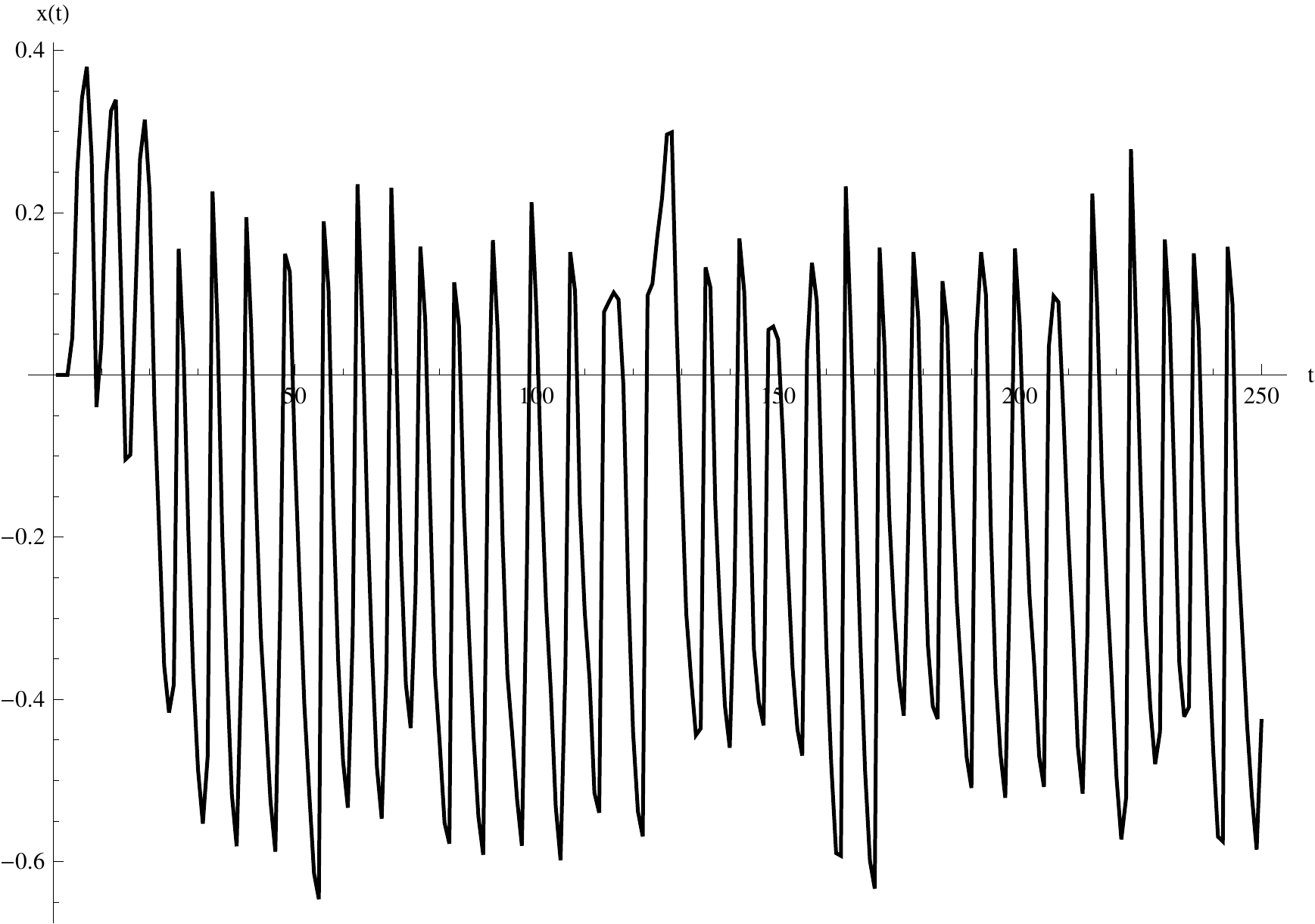}}
   \subfloat[Original series 3]{
   \includegraphics[width=0.33\textwidth]{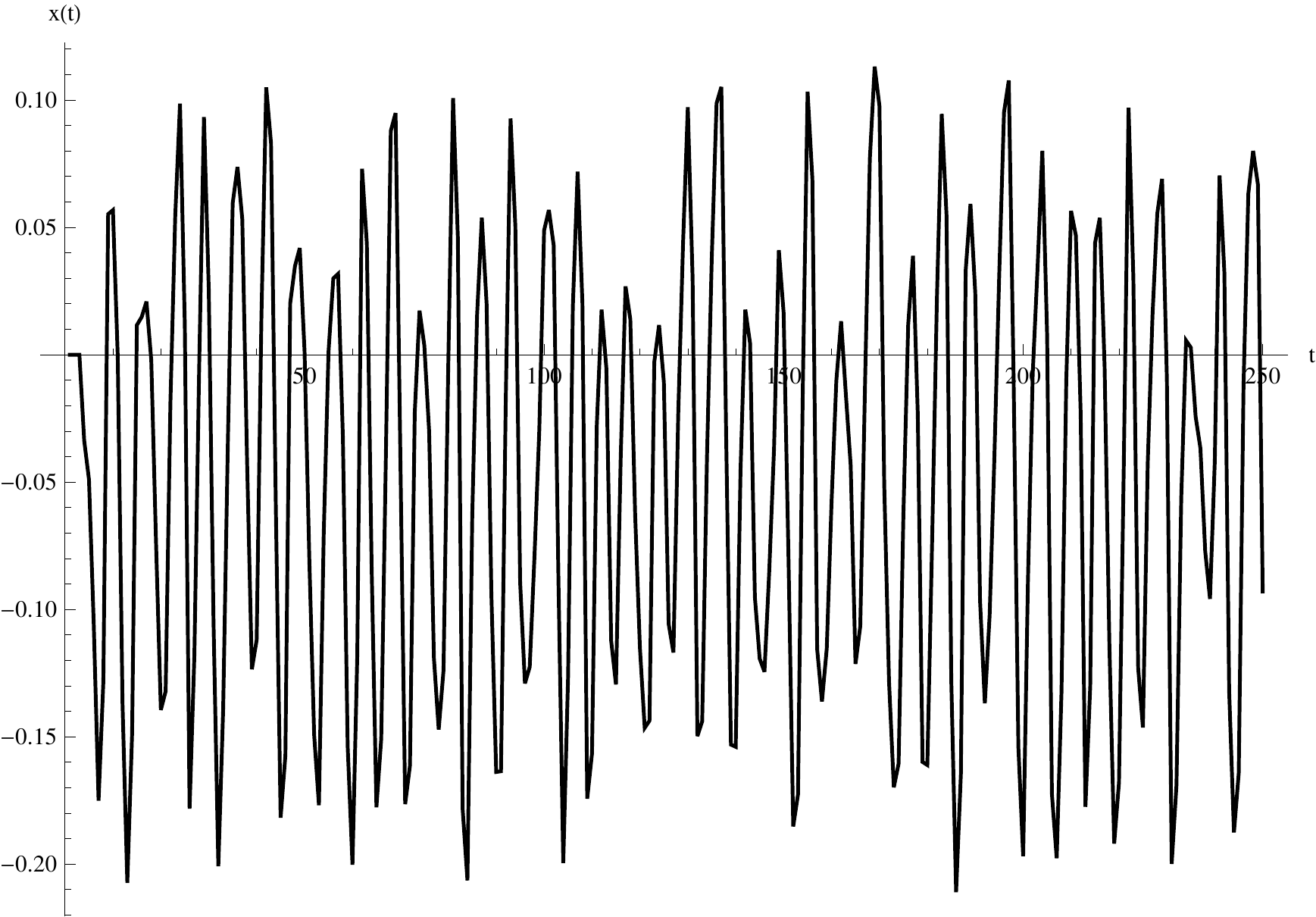}}
\\
  \subfloat[Orig. series 1 \& Herding]{
   \includegraphics[width=0.33\textwidth]{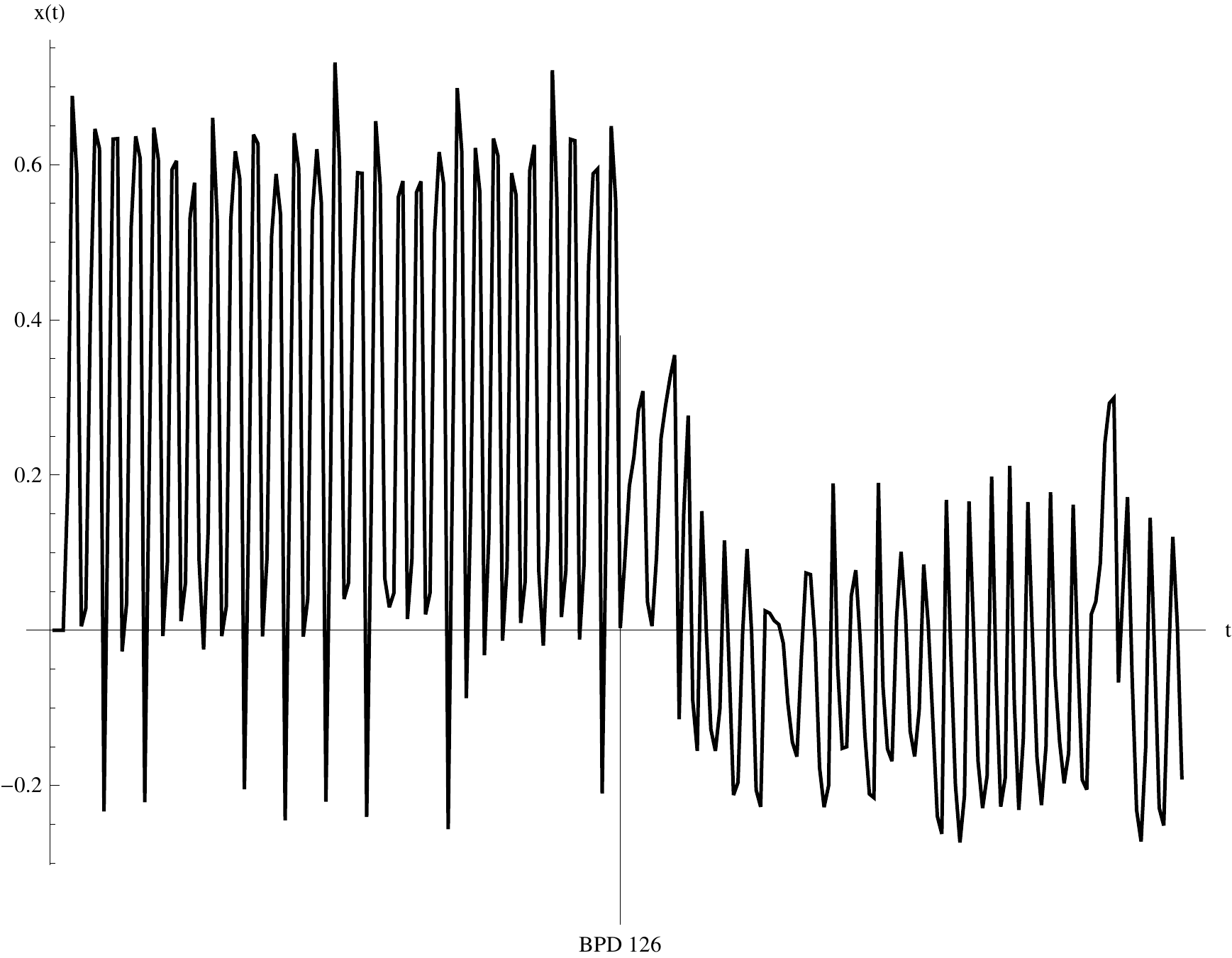}}
  \subfloat[Orig. series 2 \& Overconfidence]{
   \includegraphics[width=0.33\textwidth]{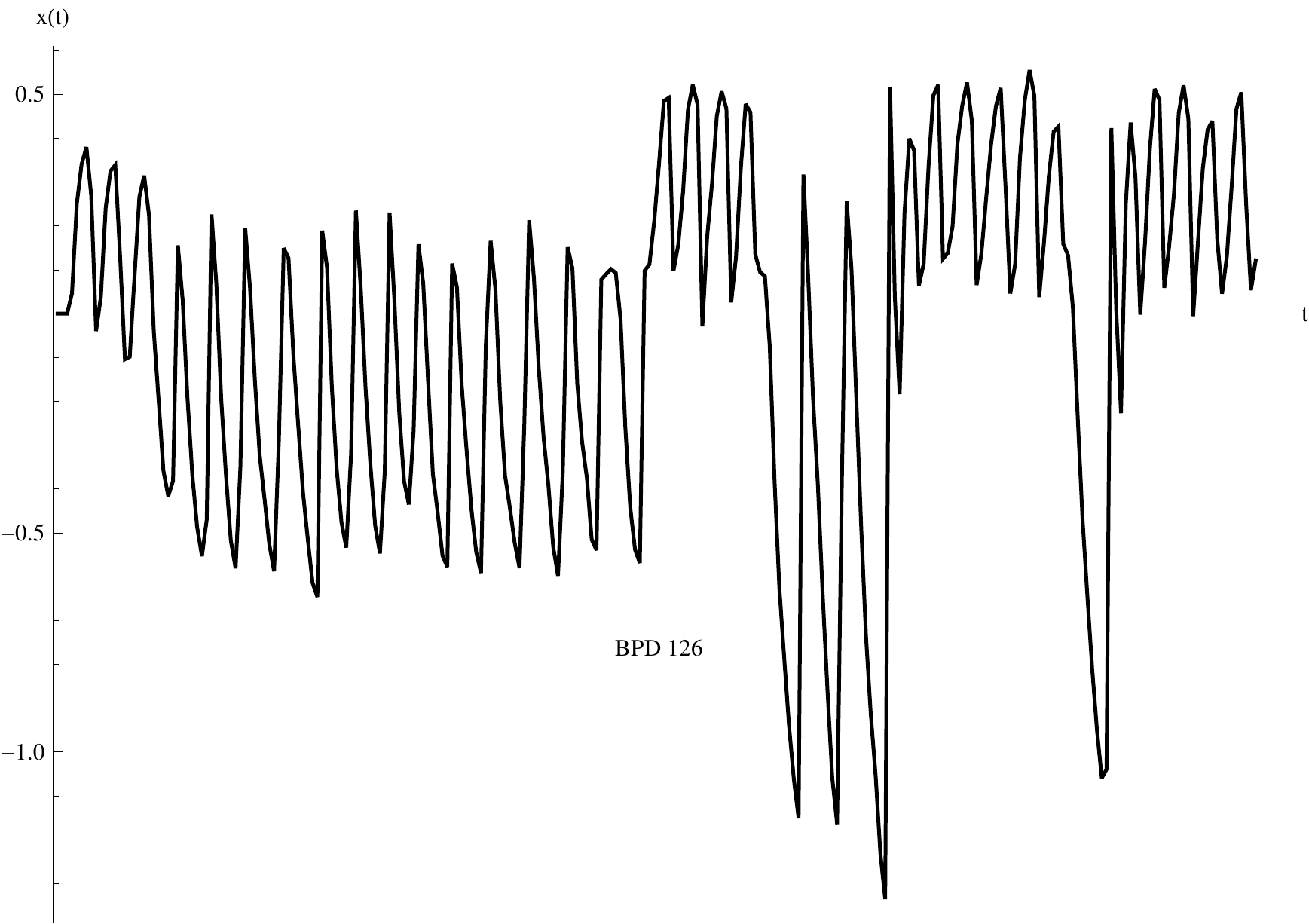}}
  \subfloat[Orig. series 3 \& Market sentiment]{
   \includegraphics[width=0.33\textwidth]{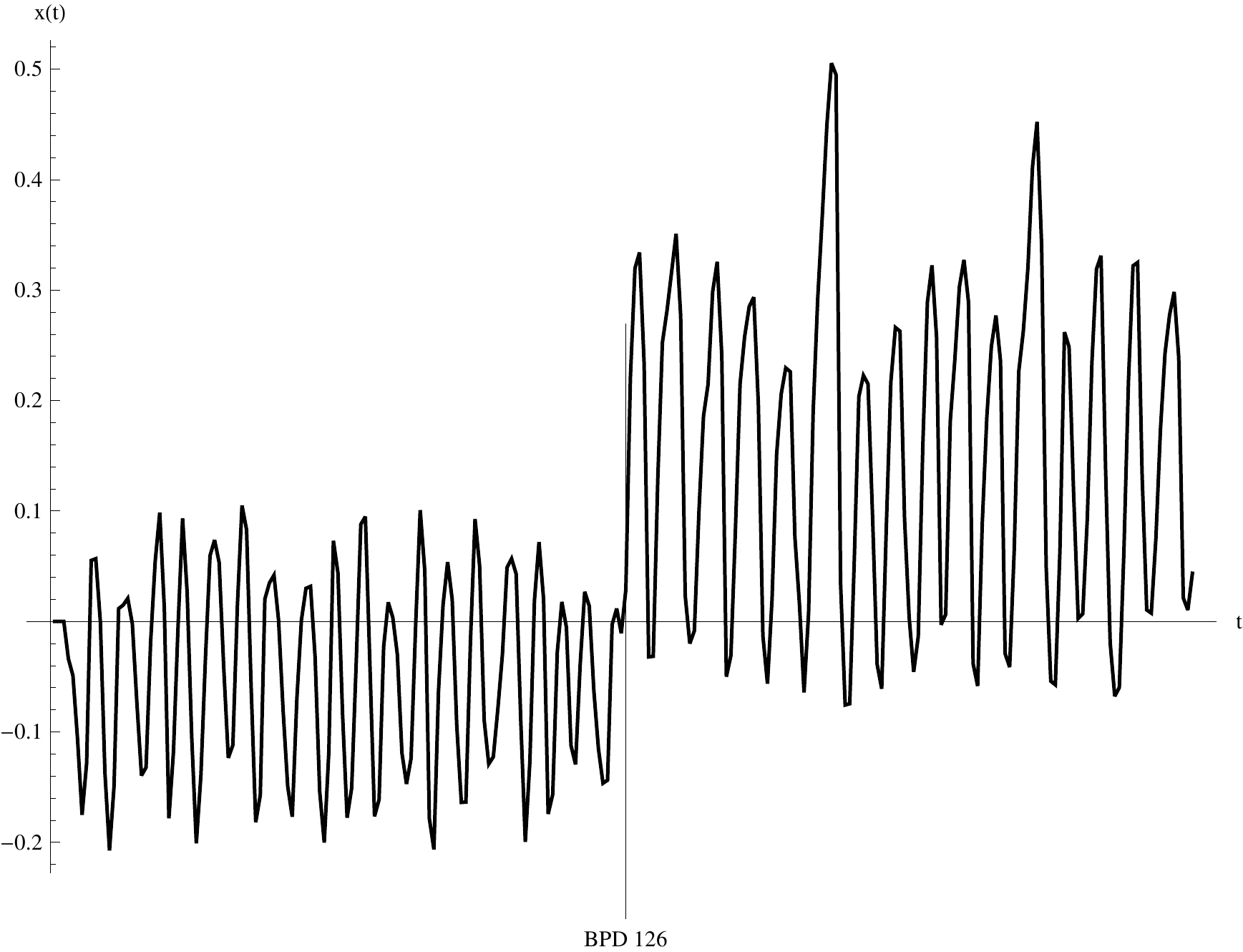}} 

\begin{center}
\end{center}
\end{figure}

To illustrate the impact of the behavioural elements incorporation on the model outcomes, \autoref{fig:typical} depicts three pairs of randomly generated series --- one for herding, one for overconfidence affecting both parameters, and one for market sentiment affecting both parameters. One can clearly see the structural change at the BPD. Regarding the single series properties, these are strongly influenced by the random combination of generated parameters and we use it here just as an illustration. However, aggregated distribution resulting from all simulations in total indeed exhibits properties which are in agreement with empirical data in financial markets.

\section{Results and interpretations}%proofred sám - 5, FINAL

In total, we simulate 13 different setup combinations including no behavioural impact, herding, three different setup combinations for overconfidence, and eight setup combinations for market sentiment according to the description in previous text. The overview of the aggregate results is summarised in \autoref{tab:results1}.

\begin{sidewaystable}[!htbp]
\caption{Simulations Results --- Overview. Note that Var. denotes variance, Skew. denotes skewness, Kurt. denotes kurtosis, +/- denote `positive/negative' market sentiment. B/A denotes `Before/After' the Break Point Date. Caps stand for the complete samples, small letters for the 20 days samples. $\varnothing \Delta$ is the arithmetic average of \% magnitude changes B/A the Break Point Date for variance and kurtosis. In the Cramér--von Mises Test for equal distributions B/A (H$_0$)  the number of non-rejections at 5\% sig. level is counted. In the Jarque--Bera Test for normality of distribution (H$_0$)  number of non-rejections at 5\% signif. level is counted.}
\label{tab:results1}
\centering
\begin{tabular*}{1\textwidth}{@{\extracolsep{\fill}}lrrrrrrrrrrrrrc} %pìkné zarovnání
\hline
\noalign{\smallskip}
~~Sample & Mean \textuparrow & Var. \textuparrow & \multicolumn{1}{c}{$\varnothing \Delta$} & Skew. \textuparrow & Kurt. \textdownarrow & \multicolumn{1}{c}{$\varnothing \Delta$} & \multicolumn{4}{c}{Cram\'er--von Mises T.} & \multicolumn{4}{c}{Jarque--Bera T.}
\\
 & \multicolumn{2}{c}{(out of 100 runs)} & \multicolumn{1}{c}{(\%)} & \multicolumn{2}{c}{(out of 100 runs)} & \multicolumn{1}{c}{(\%)} & \multicolumn{1}{c}{B-b} & \multicolumn{1}{c}{b-a} & \multicolumn{1}{c}{a-A} & \multicolumn{1}{c}{A-B} & \multicolumn{1}{c}{B} & \multicolumn{1}{c}{b} & \multicolumn{1}{c}{a} & \multicolumn{1}{c}{A} 
\\
\hline
\noalign{\smallskip}
~~No Behavioural Impact       & 48 & 53 & 9.3 & 50 & 48 & 18.5 & 100 & 100 & 100 & 100 & 2 & 7 & 7 & 2\\
~~Herding                     & 78 & 39 & 24.0 & 60 & 44 & 44.0 & 100 & 61 & 100 & 4 & 1 & 1 & 3 & 1\\
~~Overconfidence (bias)       & 47 & 97 & 81.8 & 51 & 54 & 41.7 & 100 & 18 & 100 & 2 & 1 & 4 & 7 & 0\\
~~Overconfidence (trend)      & 52 & 94 & 145.9 & 48 & 8 & 557.1 & 100 & 98 & 100 & 58 & 1 & 7 & 0 & 0\\
~~Overconfidence              & 53 & 99 & 548.7 & 53 & 4 & 695.5 & 100 & 14 & 100 & 1 & 4 & 13 & 1 & 0\\
~~M. Sentiment+ (bias)        & 100 & 79 & 33.6 & 96 & 36 & 209.9 & 100 & 3 & 100 & 0 & 1 & 8 & 5 & 2\\
~~M. Sentiment+ (trend)       & 47 & 99 & 251.9 & 47 & 9 & 641.7 & 100 & 46 & 100 & 17 & 1 & 16 & 2 & 0\\
~~M. Sentiment+ (mix)         & 100 & 3 & -20.0 & 51 & 74 & -20.7 & 100 & 1 & 100 & 0 & 1 & 6 & 16 & 4\\
~~M. Sentiment+               & 100 & 99 & 443.3 & 86 & 10 & 740.4 & 100 & 4 & 100 & 0 & 3 & 7 & 0 & 0\\
~~M. Sentiment- (bias)        & 2 & 75 & 32.9 & 10 & 40 & 56 & 100 & 3 & 100 & 0 & 1 & 6 & 6 & 0\\
~~M. Sentiment- (trend)       & 44 & 2 & -25.5 & 46 & 86 & -24.5 & 100 & 70 & 100 & 22 & 2 & 3 & 18 & 5\\
~~M. Sentiment- (mix)         & 0 & 97 & 429.1 & 18 & 13 & 713.4 & 100 & 1 & 100 & 0 & 2 & 9 & 2 & 2\\
~~M. Sentiment-               & 0 & 2 & -21.6 & 32 & 78 & -20.7 & 100 & 3 & 100 & 0 & 1 & 3 & 10 & 3 \\
\bottomrule
    \end{tabular*}
\begin{center}
\end{center}
\end{sidewaystable}

Within each simulation, we keep tracking many features. First, we evaluate the same pattern which has been revealed within the empirical benchmark sample, i.e. the shifts of mean, variance, skewness, and kurtosis between the `before' and `after' periods. While studying the empirical data, we have found that the first three of these four descriptive statistics increase in four of five analysed periods, while kurtosis decreases in the same ratio of cases. We also mention the arithmetic average of the percentual magnitude changes before and after the BPD for variance and kurtosis. 

Second, using the Cramér--von Mises Test for equal distributions we observe whether there are statistically significant differences among particular samples, namely we compare: the complete `before' sample ($B$) to the `20 day before' sample ($b$), the `20 day before' sample ($b$) to the `20 day after' sample ($a$), the `20 day after' sample (a) to the complete `after' sample (A), and finally the complete `after' sample ($A$) to the complete `before' sample ($A$). We expect $B-b$ and $A-a$ to be largely similar and, on the other hand, $b-a$ and $A-B$ to exhibit strong dissimilarities caused by the behavioural element injected into the simulation.

Third, using the \emph{Jarque--Bera Test} for normality of distribution we examine the non-normality of particular samples. We expect samples to be non-normal and strongly leptokurtic, some exceptions may appear within the 20 days samples because of only a limited number of observations after each run. For each descriptive statistic we compute the arithmetic average value. %In some cases, the results in \autoref{tab:results1} and \autoref{tab:results2} might appear at odds that this is caused by different methodology of collecting results --- in \autoref{tab:results1} we just count how many setup combinations out of total 100 fulfill particular feature regardless their magnitude while figures in \autoref{tab:results2} depicting the arithmetic averages are more vulnerable to be influenced by several markedly small or large numbers. We always depict values for the `20 days before' sample and `20 day after' sample and for variance and kurtosis we compute the percentual change in the average magnitude before and after the BPD. Finally, we depict maximum and minimum values and generally expect higher fluctuations in the `after' samples. 

\subsection{Simulations results}
\label{sec:res}

According to the Jarque--Bera Test for normality of distribution, simulated samples are largely non-normal with substantial excess kurtosis, thus likely to exhibit leptokurtic properties and confirm our expectation. This fundamental finding is thus consistent with the empirical benchmark sample where all subsamples exhibit excess kurtosis. For the complete samples, there is only several cases of normality, perhaps those with low intensity of behavioural element. However, several 20 day samples --- especially the market sentiment cases --- reach considerable values. It seems that market sentiment affecting the trend $g_h$ only and the `mixed' case (also trend-affecting) produce samples closer to normal distribution than all other setup combinations. Apparently, modifications affecting trend $g_h$ when the behavioural `sentiment element' is included have certain tendency to offset the ability to produce real market-like leptokurtic distributions --- one of the most highlighted features of the model. This finding is, however, partially consistent with the benchmark sample tendency to exhibit decreases in kurtosis after the BPD.

\subsection{Empirical pattern fitting} 

Let us look on results of the three behavioural modifications of the model. Herding seems to affect the model structure the least. It exhibits more or less an average effect on all descriptive statistics with prevailing effect of kurtosis increase which goes counter the empirical findings. Although the presence of the herding towards the most profitable strategy produces some minor differences, namely mean shifts and dissimilarities in distributions, herding effect is comparatively rather similar to the case without any behavioural impact. When comparing the distributions of the data 20 days before and after the herding occurs in the model ($b-a$ case) we can see that distribution does not change significantly in 61 out of 100 cases.
	
While herding does not bring any strong results, overconfidence seems to influence the model much more significantly. All three overconfidence setups increase variance in the large majority of cases (94 to 99 out of 100). This result is comparable with our empirical findings as well as  with conclusions of \citet{DaHiSu1998} and in fact is expected. On the other hand, overconfidence does not affect the mean of the distributions. An intriguing feature is the rapid variance and kurtosis increase after the BPD in both cases with the trend overconfidence. When comparing these results to the findings from the empirical data we can see that overconfidence affecting the bias parameter $b_h$ only reveals substantially higher ability to fit the decreasing kurtosis empirical pattern. At the same time it does not produce such extensive variance and especially kurtosis increases. When thinking about an economic interpretation, one can understand this specific setup as a situation when all market participant strictly use similar pricing models and thus their trend extrapolation is not a subject to any bias, while their personal feelings and expectations of the market future development are highly impacted by overconfidence. Moreover, only in 18 out of 100 cases we can not distinguish between the distributions of generated data 'before' and 'after' the BPD. In the case of overconfidence affecting trend parameter, 98 out of 100 cases can not be distinguished from each other. This means that overconfidence affecting trend does not change the distribution of generated data at all. Thus we conclude that the case of overconfidence affecting bias matches the empirical data best.

Turning our attention to the market sentiment we find out that it seems to be --- compared to the previous modifications --- the most promising behavioural change of the model structure. At first glance, effects of the positive and the negative market sentiment generate roughly inverse results. The positive market sentiment fits the empirical benchmark sample pattern considerably better, whereas the negative market sentiment is able only to mimic decreases in kurtosis (2 of 4 cases) and variance increases (2 of 4 cases). This might be explained in a similar way as we offer for the positive mean shifts within the empirical benchmark sample.
	
As the positive sentiment matches the behaviour found in the empirical data best, we focus on its results mainly. Market sentiment affecting trend $g_h$ seems to be a weak modification as it exhibits average values for the mean and skewness shifts and has very low performance in the case of kurtosis. Mixed sentiment cancels the important variance shift almost entirely out but is able to mimic the kurtosis decrease. Again, we observe excessive variance and especially kurtosis upward jumps when the trend affecting sentiment is introduced. Sentiment changes affecting either both trend $g_h$ and bias $b_h$ or bias $b_h$ only seem to be the most successful modifications and we again conclude that the market sentiment affecting bias parameter $b_h$ only best matches the empirical findings. This particular modification embodies higher performance in the kurtosis decrease fitting and, most importantly, it exhibits much more reasonable percentual changes of variance and kurtosis in comparison with the empirical values.

Employing the Cram\'er--von Mises Test for equal distributions we study the ability of particular behavioural modifications to produce significantly different data distributions before and after the BPD. From this perspective, herding and all trend affecting setup combinations generally seem to be the weakest modifications. On the other hand, almost all non-trend sentiment modification exhibit excellent results from this point of view. Our expectation that the 20 day samples and the complete samples from the same period come from the same distribution has been 100\% affirmed as no single deviation appeared.

Increases of variance seem to be the most robust results of the majority of the overconfidence and market sentiment setup combinations. This finding confirms the conclusions of \citet{DaHiSu1998} and \citet{DiWe2005}. The tendency of the minimum and maximum values then also more or less confirms our expectation of higher volatility after the BPD. Nonetheless, the minimum and maximum values are rather chance results as they only represent a single extreme value of all 100 runs.

%***********************

\subsection{Model extensions}
\label{sec:ext}

While being rather successful in replicating empirical findings by behavioural injections in the HAM model, we focus on further possibilities of model extensions in order to test robustness of our results. In further analysis, we drop the setups with negative market sentiment as these are far from the empirical data.

The overviews of the aggregate results are summarised in \autoref{tab:results3} \& \ref{tab:results3b}. The structure of the tracked information remains the same as in the previous analysis.

\begin{sidewaystable}[!htbp]
\caption{Extensions results I. --- overview. Note that Var. denotes variance, Skew. denotes skewness, Kurt. denotes kurtosis, +/- denote `positive/negative' market sentiment. B/A denotes `Before/After' the Break Point Date. Caps stand for the complete samples, small letters for the 20 days samples. $\varnothing \Delta$ is the arithmetic average of \% magnitude changes B/A the Break Point Date for variance and kurtosis. In the Cramér--von Mises Test for equal distributions B/A (H$_0$)  the number of non-rejections at 5\% sig. level is counted. In the Jarque--Bera Test for normality of distribution (H$_0$)  number of non-rejections at 5\% signif. level is counted.}

\label{tab:results3}
\centering
\begin{tabular*}{0.96\textwidth}{@{\extracolsep{\fill}}lrrrrrrrrrrrrrc} %pìkné zarovnání
\hline
\noalign{\smallskip}
~~Extension \& Sample & Mean \textuparrow & Var. \textuparrow & \multicolumn{1}{c}{$\varnothing \Delta$} & Skew. \textuparrow & Kurt. \textdownarrow & \multicolumn{1}{c}{$\varnothing \Delta$} & \multicolumn{4}{c}{Cram\'er--von Mises T.} & \multicolumn{4}{c}{Jarque--Bera T.}
\\
 & \multicolumn{2}{c}{(out of 100 runs)} & \multicolumn{1}{c}{(\%)} & \multicolumn{2}{c}{(out of 100 runs)} & \multicolumn{1}{c}{(\%)} & \multicolumn{1}{c}{B-b} & \multicolumn{1}{c}{b-a} & \multicolumn{1}{c}{a-A} & \multicolumn{1}{c}{A-B} & \multicolumn{1}{c}{B} & \multicolumn{1}{c}{b} & \multicolumn{1}{c}{a} & \multicolumn{1}{c}{A}
 \\
\hline
\noalign{\smallskip}
~~Fundamentalists:            &  &  &  &  &  &  &  &  &  &  &  &  &  &
\\
\noalign{\medskip}
~~Herding                     & 76 & 29 & 14.0 & 73 & 37 & 30.2 & 100 & 38 & 100 & 1 & 0 & 1 & 0 & 0
\\
~~Overconfidence (bias)       & 51 & 99 & 86.3 & 46 & 52 & 57.5 & 100 & 19 & 100 & 4 & 1 & 6 & 5 & 0
\\
~~Overconfidence (trend)      & 45 & 98 & 127.8 & 42 & 6 & 419.8 & 100 & 99 & 100 & 70 & 2 & 9 & 2 & 1
\\
~~Overconfidence              & 52 & 98 & 103.8 & 53 & 66 & 50.1 & 100 & 20 & 100 & 4 & 0 & 4 & 2 & 1
\\
~~M. Sentiment+ (bias)        & 99 & 78 & 32.9 & 93 & 32 & 107.4 & 100 & 3 & 100 & 0 & 2 & 3 & 3 & 2
\\
~~M. Sentiment+ (trend)       & 56 & 98 & 241.7 & 60 & 6 & 524.7 & 100 & 49 & 100 & 21 & 0 & 9 & 0 & 0
\\
~~M. Sentiment+ (mix)         & 100 & 2 & -27.8 & 65 & 84 & -27.0 & 100 & 2 & 100 & 0 & 0 & 2 & 14 & 2
\\
~~M. Sentiment+               & 100 & 98 & 394.4 & 87 & 8 & 606.0 & 100 & 3 & 100 & 0 & 1 & 3 & 1 & 0
\\
\noalign{\medskip}
~~Stochastic Parameters:      &  &  &  &  &  &  &  &  &  &  &  &  &  &
\\
\noalign{\medskip}
~~Herding                     & 80 & 36 & -5.1 & 47 & 50 & 56.5 & 100 & 76 & 100 & 13 & 1 & 7 & 7 & 1
\\
~~Overconfidence (bias)       & 41 & 98 & 74.8 & 52 & 51 & 27.2 & 100 & 0 & 100 & 0 & 2 & 6 & 5 & 1
\\
~~Overconfidence (trend)      & 50 & 99 & 167.8 & 52 & 2 & 765.8 & 100 & 100 & 100 & 54 & 3 & 11 & 0 & 0
\\
~~Overconfidence              & 58 & 100 & 317.1 & 56 & 3 & 636.4 & 100 & 0 & 100 & 0 & 4 & 8 & 1 & 0
\\
~~M. Sentiment+ (bias)        & 100 & 86 & 20.7 & 94 & 37 & 61.4 & 100 & 0 & 100 & 0 & 2 & 7 & 1 & 0
\\
~~M. Sentiment+ (trend)       & 51 & 99 & 316.0 & 54 & 1 & 957.8 & 100 & 18 & 100 & 0 & 0 & 4 & 0 & 0
\\
~~M. Sentiment+ (mix)         & 100 & 3 & -21.7 & 44 & 89 & -22.8 & 100 & 0 & 100 & 0 & 0 & 6 & 24 & 5
\\
~~M. Sentiment+               & 100 & 100 & 356.4 & 98 & 3 & 946.0 & 100 & 0 & 100 & 0 & 3 & 5 & 0 & 0
\\
\bottomrule
    \end{tabular*}
\begin{center}
\end{center}
\end{sidewaystable}

\begin{sidewaystable}[!htbp]
\caption{Extensions Results II. --- Overview. Note that Var. denotes variance, Skew. denotes skewness, Kurt. denotes kurtosis, +/- denote `positive/negative' market sentiment. B/A denotes `Before/After' the Break Point Date. Caps stand for the complete samples, small letters for the 20 days samples. $\varnothing \Delta$ is the arithmetic average of \% magnitude changes B/A the Break Point Date for variance and kurtosis. In the Cramér--von Mises Test for equal distributions B/A (H$_0$)  the number of non-rejections at 5\% sig. level is counted. In the Jarque--Bera Test for normality of distribution (H$_0$)  number of non-rejections at 5\% signif. level is counted.}
\label{tab:results3b}
\centering
\begin{tabular*}{0.96\textwidth}{@{\extracolsep{\fill}}lrrrrrrrrrrrrrc} %pìkné zarovnání
\hline
\noalign{\smallskip}
~~Extension \& Sample & Mean \textuparrow & Var. \textuparrow & \multicolumn{1}{c}{$\varnothing \Delta$} & Skew. \textuparrow & Kurt. \textdownarrow & \multicolumn{1}{c}{$\varnothing \Delta$} & \multicolumn{4}{c}{Cram\'er--von Mises T.} & \multicolumn{4}{c}{Jarque--Bera T.}
\\
 & \multicolumn{2}{c}{(out of 100 runs)} & \multicolumn{1}{c}{(\%)} & \multicolumn{2}{c}{(out of 100 runs)} & \multicolumn{1}{c}{(\%)} & \multicolumn{1}{c}{B-b} & \multicolumn{1}{c}{b-a} & \multicolumn{1}{c}{a-A} & \multicolumn{1}{c}{A-B} & \multicolumn{1}{c}{B} & \multicolumn{1}{c}{b} & \multicolumn{1}{c}{a} & \multicolumn{1}{c}{A}
\\
\hline
\noalign{\smallskip}
~~Combinations:               &  &  &  &  &  &  &  &  &  &  &  &  &  &
\\
\noalign{\medskip}
~~Herd. \& Over. (bias)       & 88 & 49 & 52.7 & 62 & 33 & 137.9 & 100 & 72 & 100 & 8 & 3 & 8 & 4 & 1
\\
~~Herd. \& Sent.+ (bias)      & 99 & 58 & 111.3 & 87 & 29 & 156.2 & 100 & 0 & 100 & 0 & 1 & 8 & 1 & 0
\\
~~Over. (b.) \& Sent.+ (b.)   & 100 & 97 & 112.7 & 91 & 31 & 80.8 & 95 & 1 & 100 & 0 & 1 & 10 & 3 & 2
\\
~~H. \& O. (b.) \& S.+ (b.)   & 100 & 66 & 58.6 & 86 & 33 & 122.9 & 100 & 0 & 100 & 0 & 1 & 10 & 2 & 0
\\
\noalign{\medskip}
~~Memory:                     &  &  &  &  &  &  &  &  &  &  &  &  &  &
\\
\noalign{\medskip}
~~Herding                     & 85 & 69 & 70.9 & 65 & 43 & 60.1 & 97 & 33 & 95 & 10 & 0 & 1 & 0 & 0
\\
~~Overconfidence (bias)       & 49 & 95 & 135.2 & 47 & 53 & 101.9 & 100 & 11 & 98 & 3 & 1 & 2 & 2 & 1
\\
~~Overconfidence (trend)      & 41 & 95 & 96.2 & 46 & 50 & 28.7 & 97 & 15 & 95 & 3 & 0 & 1 & 2 & 1
\\
~~Overconfidence              & 51 & 99 & 362.1 & 52 & 8 & 400.4 & 99 & 16 & 96 & 3 & 0 & 1 & 1 & 0
\\
~~M. Sentiment+ (bias)        & 99 & 64 & 122.2 & 77 & 48 & 161.3 & 100 & 2 & 37 & 0 & 0 & 4 & 2 & 0
\\
~~M. Sentiment+ (trend)       & 55 & 95 & 299.2 & 59 & 18 & 345.3 & 100 & 25 & 96 & 10 & 0 & 0 & 1 & 0
\\
~~M. Sentiment+ (mix)         & 100 & 4 & -30.1 & 67 & 76 & -15.7 & 99 & 1 & 34 & 0 & 0 & 2 & 0 & 0
\\
~~M. Sentiment+               & 98 & 97 & 489.5 & 84 & 14 & 506.6 & 98 & 1 & 31 & 0 & 1 & 2 & 0 & 0
\\
\bottomrule
    \end{tabular*}
\begin{center}
\end{center}
\end{sidewaystable}

%***********************

\subsubsection{Fundamentalists by default}
%\label{sec:funddef}

Fundamentalists play crucial role in the theoretical model framework. Until now, we allowed a fundamental strategy to be generated only within random setting, but next we would like to study the change of the market behaviour in case when the one fundamental strategy will surely be present in the market. Therefore, we will always set $g_1=b_1=0$ for one of strategies.

Looking at the results, fundamentalists' default presence seems to have only a minor and non-systematic impact on the model outcomes. It affects mainly kurtosis and moves only two setup combinations slightly closer to the empirical benchmark sample. The effect on kurtosis is also evident from the results of the Jarque--Bera Test which rejects normality in more cases than within the primary results.

In the overconfidence case, fundamentalists strongly decrease variance but only for the setup combination where both trend $g_h$ and bias $b_h$ are affected. For the same setup kurtosis strongly increases in the `before' sample and strongly decreases in the `after' sample. In this specific case fundamentalists thus balance the kurtosis magnitudes before and after the BPD.

In the market sentiment case, we can also observe decreasing fundamentalists' impact on kurtosis, but only in the `mix' case.

%***********************

\subsubsection{Stochastic formation of parameters}
\label{sec:stochform}

It is perhaps more realistic to assume that the levels of the market sentiment or overconfidence as well as the tendency to switch between particular strategies vary both among traders and in time. In an effort to reflect this sensible idea, we introduce the stochastic formation of the intensity of choice $\beta$ as well as the behavioural element intensity. Both parameters are newly generated randomly for each simulation, which means we employ 10.000 different random combinations of these two model inputs for each setup combinations. Parameter ranges remain the same, i.e.:

\begin{itemize}
	\item The intensity of choice $\beta$ is drawn from the uniform distribution $U(5,500)$;
	\item The intensity of overconfidence is drawn from the uniform distribution $U(0.05,0.5)$;
  \item The intensity of the market sentiment is drawn from the uniform distribution $U(0.04,0.4)$ for trend $g_h$ and from $U(0.03,0.3)$ for bias $b_h$.
\end{itemize}

\autoref{tab:results3} reveals that stochastic formation of parameters does not bring any considerable difference to model outcomes. In fact, this modification does not change the structure of the results, so our previous findings are robust to change of parameters in the model.

%***********************

\subsubsection{Combinations}
\label{sec:combi}

Next, we consider the possible combinations of more behavioural modifications in the model and examine whether combinations of impacts lead rather to a consolidation of results or to canceling the effects out. We combine four setups which best match the empirical findings. Hence, following combinations emerge:

\begin{itemize}
  \item Herding \& overconfidence affecting the bias parameter $b_h$ only;
  \item Herding \& market sentiment affecting the bias parameter $b_h$ only;
  \item Overconfidence \& market sentiment affecting the bias parameter $b_h$ only;
  \item Combination of all these three modifications.
\end{itemize}

Results of these simulations seem ambiguous and at first glance it is not clear which component dominates. The combinations of various modifications lead more likely to canceling the effects out than to a consolidation of results. Market sentiment seems to be the governing element, but its effect is comparatively smaller when it is combined with other modifications than when it works alone. Herding follows and outperforms the impact of overconfidence. Compared with the primary findings, no strong tendencies are apparent, new results are most often either the same or somewhere `half-way' between original results.
 
%***********************

\subsubsection{Memory}
\label{sec:mem}

The effect of memory is undoubtedly one of the most interesting features which can be studied within the heterogeneous agent modelling. The real-life parallel with the human memory and learning is apparent. As \citet[pg. 56]{Hommes2006} points out, \emph{``another important issue is how memory in the fitness measure affects stability of evolutionary adaptive systems and survival of technical trading''}. Memory effect has been widely examined in various studies and we also incorporate this notion into our system. Memory process is the most distinctive modification of the model structure from all four considered extensions and for its implementation we use the similar approach as \citet{BaVaVo2009} and \citet{VaBaVo2009}. More formally, \autoref{eq:setting4} is extended to the form:
\begin{eqnarray}
\label{eq:settingmem}
U_{h,t-1}&=&\frac{1}{m_h}\sum_{l=0}^{m_h-1}\left[(x_{t-1-l}-Rx_{t-2-l})\frac{\mathnormal{f}_{h,t-2-l}-Rx_{t-2-l}}{a\sigma^2}\right]\nonumber \\
&\equiv& \frac{1}{m_h}\sum_{l=0}^{m_h-1}\left[(x_{t-1-l}-Rx_{t-2-l})\frac{g_h x_{t-3-l}+b_h-Rx_{t-2-l}}{a\sigma^2}\right],
\end{eqnarray}
where $m_h$ denotes the memory length of each particular strategy $h$. In simulations, memory lengths $m_h$ are generated randomly from the uniform distribution $U(0,20)$. This range has been defined with regard to computational time necessary for the memory implementation (which is in this particular setting more than 2.5 times higher than with zero memory) and also considering the length of the observed periods before and after the BPD.

Although memory has surely the most evident impact on the model outcomes, it still does not cause major dissimilarities compared to our primary results. The presence of memory impacts mainly the structure of data distributions and make them substantially mutually distinct and non-normal as is evident from the aggregate results of the Cramér--von Mises and Jarque--Bera Tests --- the number of rejections of equal distributions rises almost for all setup combinations and the same holds for normality. Moreover, memory enormously affects the structure of the `20 day after' sample which does no longer share the same distribution with the `complete after' sample for the majority of runs. This result suggests a strong memory effect around the BPD. Memory also influences kurtosis and moves the ratio of grows and declines closer to the empirical benchmark sample values for some setup combinations. Considering the modification which best matches the findings from empirical data --- the case of the market sentiment affecting the bias $b_h$ parameter only --- memory slightly reinforces its ability to fit the benchmark sample pattern. Memory perhaps also partially affects skewness and variance, but these deviations are largely non-systematic and might be rather chance results.

\begin{figure}[h]%[!htbp]
\caption{Illustration of generated data for different behavioural breaks in the model with memory. The upper part shows the data simulated from original model, bottom part depicts counterpart of the simulation where (d) herding, (e) overconfidence, and (f) market sentiment is injected.}
\label{fig:typicalmem}
\vspace{1em}
\centering
  \subfloat[Original series 1]{
   \includegraphics[width=0.33\textwidth]{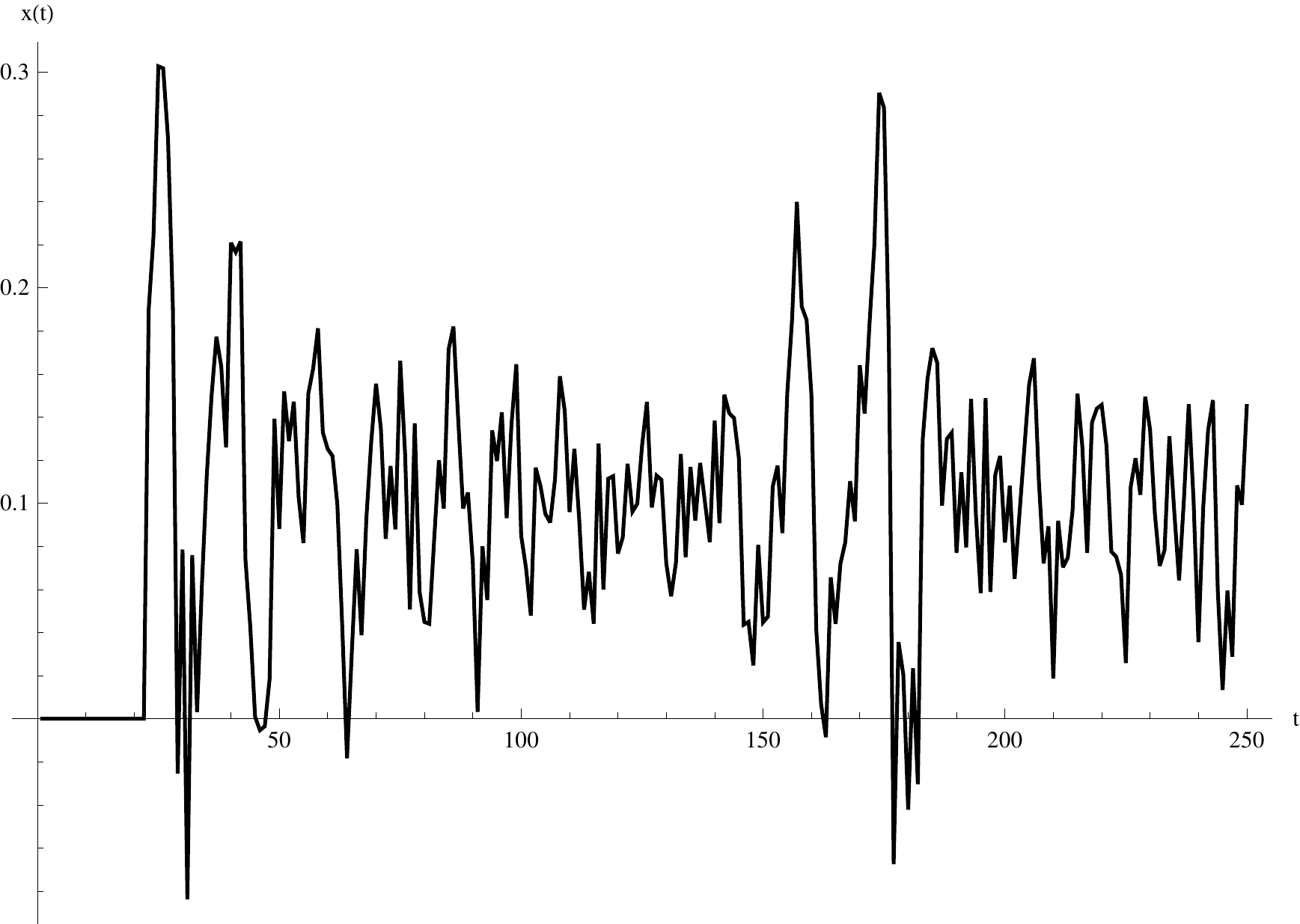}}
   \subfloat[Original series 2]{
   \includegraphics[width=0.33\textwidth]{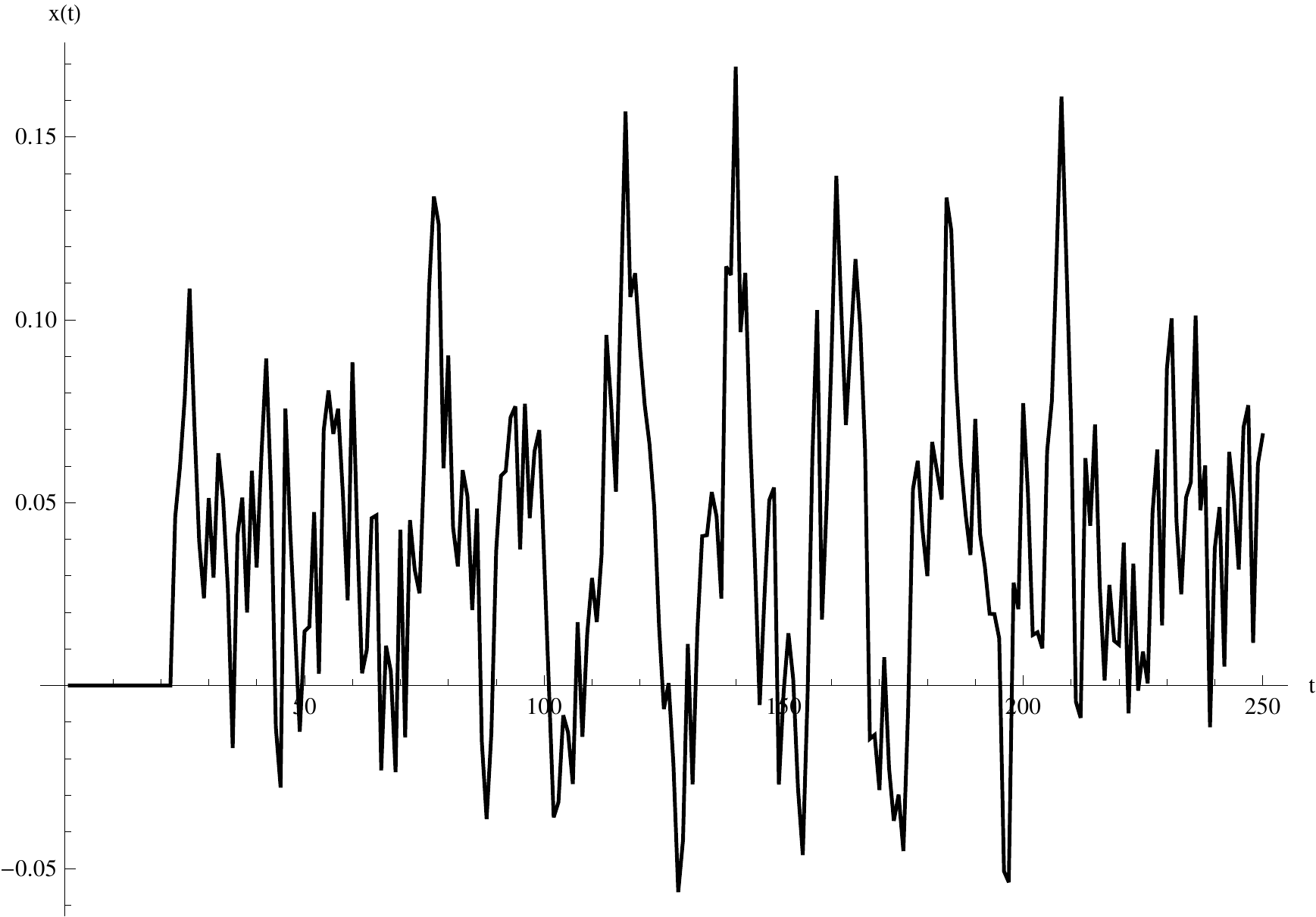}}
   \subfloat[Original series 3]{
   \includegraphics[width=0.33\textwidth]{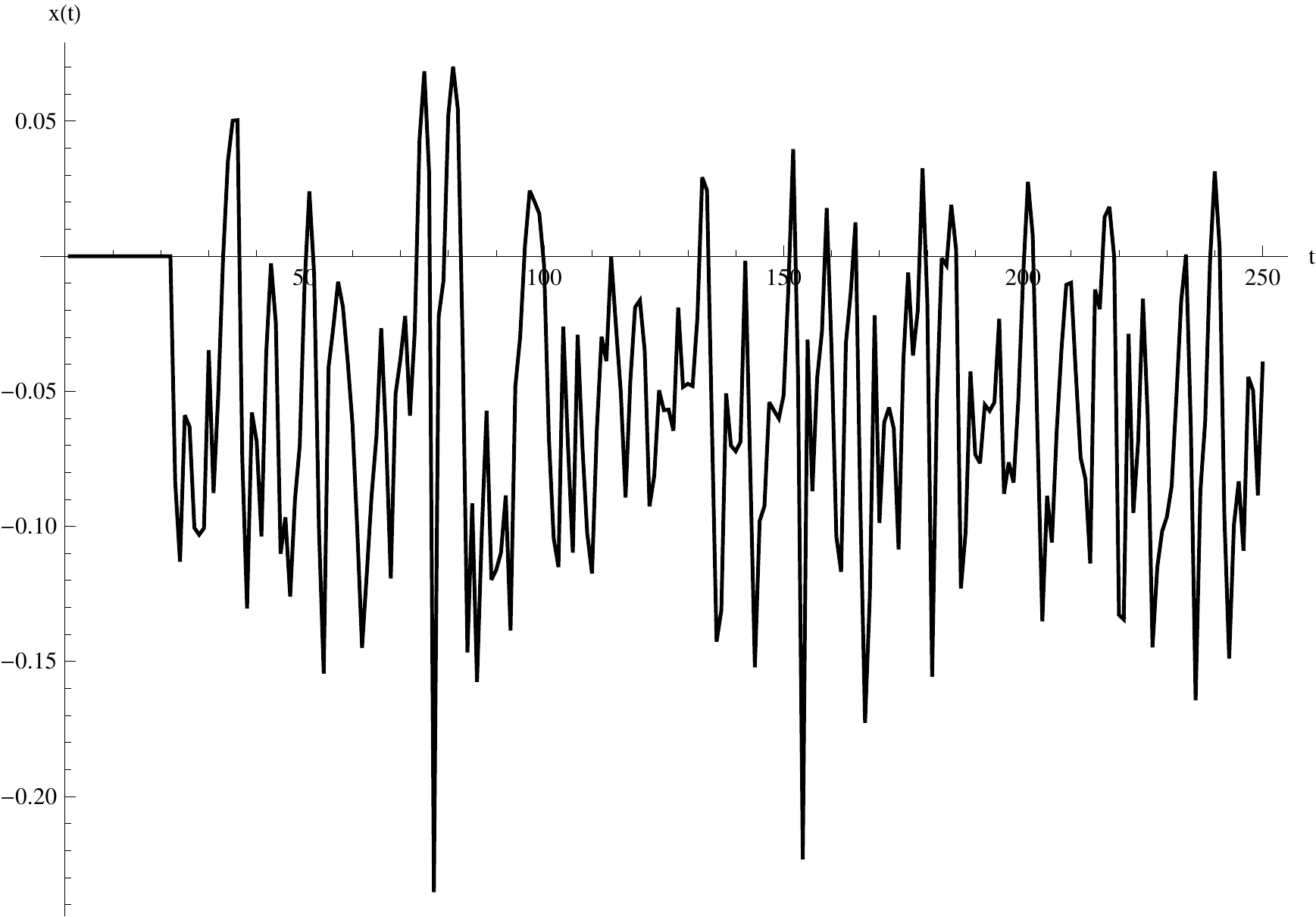}}
\\
  \subfloat[Orig. series 1 \& Herding]{
   \includegraphics[width=0.33\textwidth]{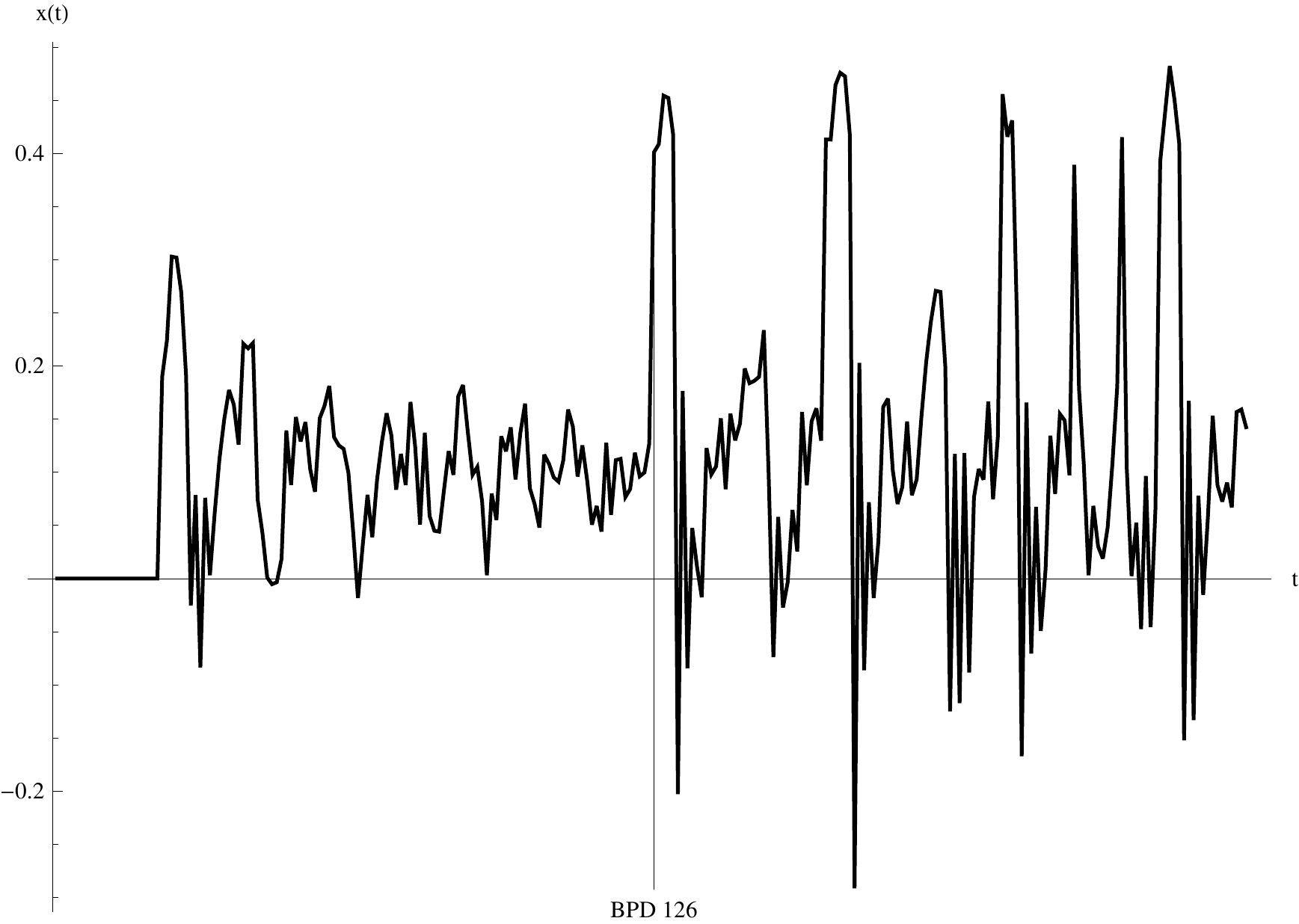}}
  \subfloat[Orig. series 2 \& Overconfidence]{
   \includegraphics[width=0.33\textwidth]{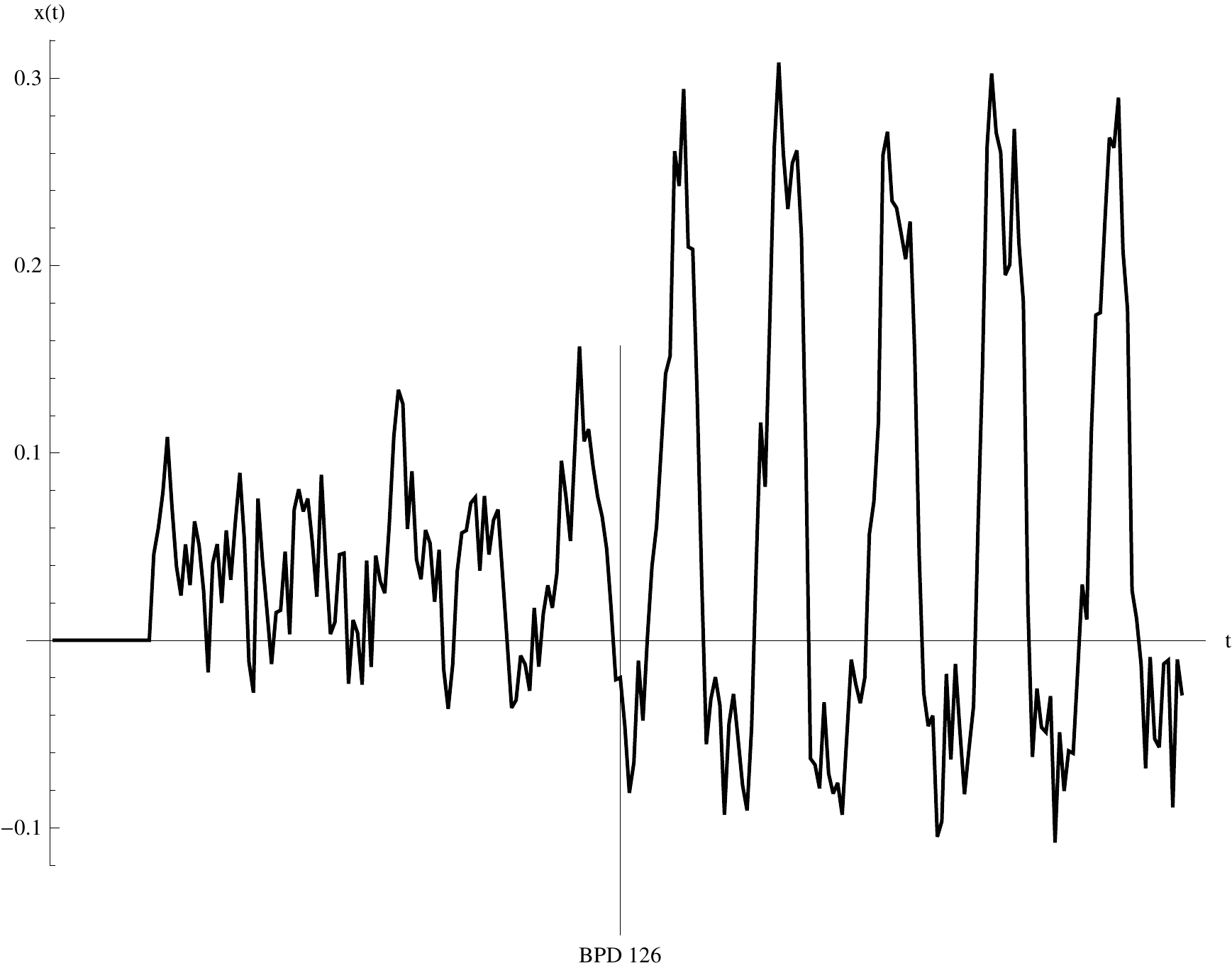}}
  \subfloat[Orig. series 3 \& Market sentiment]{
   \includegraphics[width=0.33\textwidth]{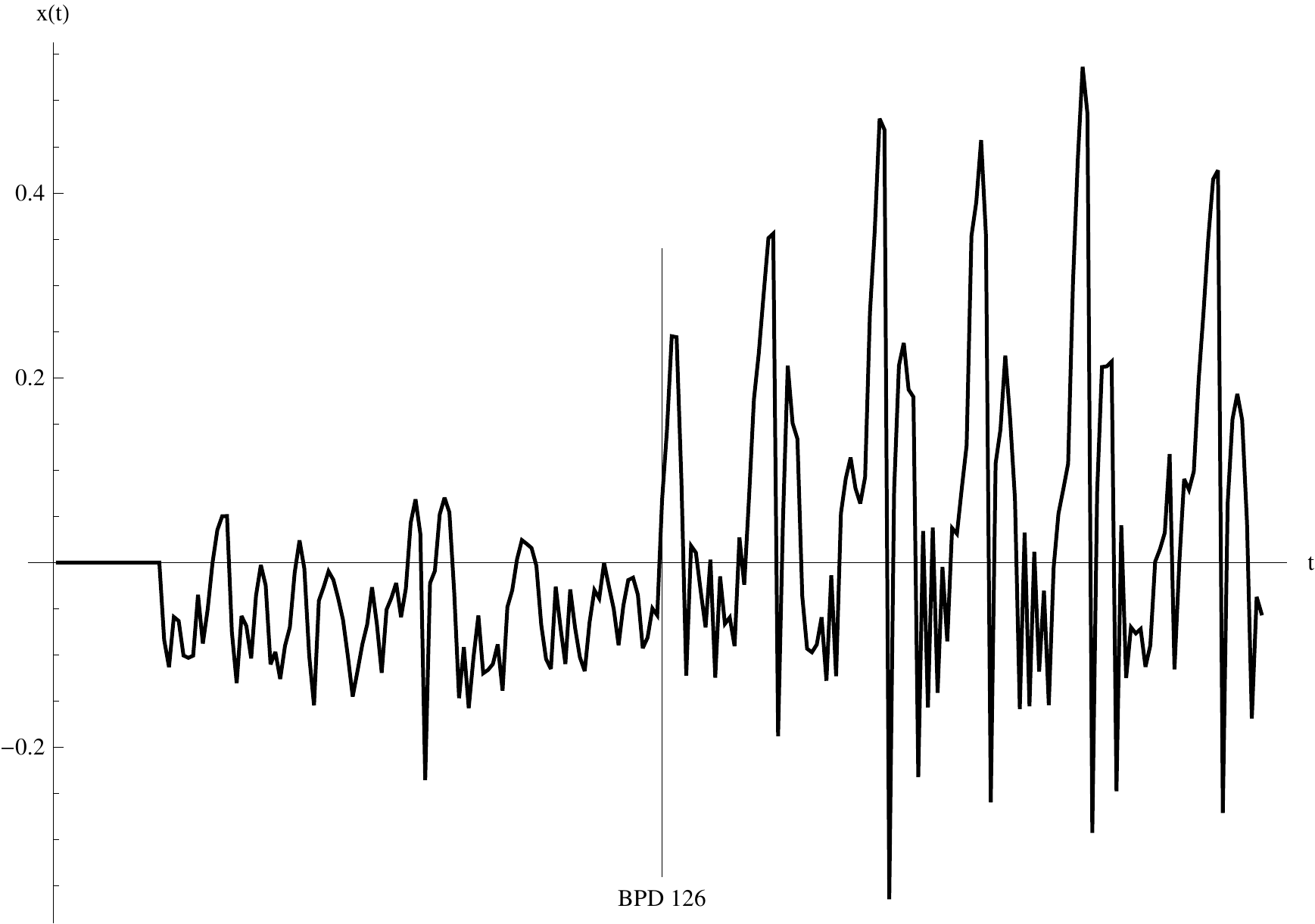}}
\begin{center}
\end{center}
\end{figure}

Again, to illustrate the impact of the behavioural elements incorporation on the model with memory outcomes, \autoref{fig:typicalmem} depicts three pairs of randomly generated series --- one for herding, one for overconfidence affecting both parameters, and one for market sentiment affecting both parameters. One can again clearly see the structural change at the BPD.

%***********************

\section{Conclusion}
\label{sec:igr}

We introduce a different perspective and application of the \citet{BrHo1998} HAM approach. From the plethora of well documented behavioural biases we examine the impact of herding, overconfidence, and market sentiment. Behavioural patterns are embedded into an asset pricing framework in order to study resulting price dynamics.

As financial crises and stock market crashes can be widely considered as periods when investors' rationality is restrained to a great extent, we advance current research literature through an empirical verification of the HAM abilities and explanatory power of behavioural finance, using data covering these periods of high-volatility. We develop a unique benchmark dataset for this purpose. The dataset consists of all currently publicly accessible individual DJIA constituents covering five particularly turbulent stock market periods. The era we consider starts with Black Monday 1987 and terminates with the Lehman Brothers bankruptcy in 2008. Most importantly, we reveal an interesting pattern among this data. When data before and after the BPD are compared, mean, variance, and skewness increase in four of five analysed periods, while kurtosis decreases in the same ratio of cases.

From the theoretical point of view, we show that selected findings from behavioural finance:  (i) herding, (ii) overconfidence, and (iii) market sentiment, can be well modelled via the HAM. Herding is modelled so that one of strategies copies the behaviour of the most successful traders of the previous day. Overconfidence is modelled as the overestimation of randomly generated values. Finally, we model the market sentiment as shifts of the mean values of probability distributions from which agents' beliefs are generated. 

A numerical analysis of the HAM extended with the selected findings from the behavioural finance brings the most important results. We analyse the dynamics of the model around the BPD, where behavioural elements are injected into the system, and compare it to our empirical benchmark sample. By performing large Monte Carlo study we show that findings from behavioural finance extend the original HAM considerably and different HAM modifications lead to different outcomes. Both results are statistically significant and are confirmed by the Cramér--von Mises Test. From this perspective, herding and all trend-affecting setup combinations generally seem to be the weakest modifications. On the other hand, almost all non-trend sentiment modifications exhibit excellent results. Moreover, we analyse the impact of four additional model extensions. These comprise the default presence of fundamentalists, stochastic formation of parameters, various different combinations, and memory. Generally, the first three modifications cause more or less only minor differences, while memory exhibits the most evident and statistically significant influence on the model structure. Results further indicate that HAM is able to partially replicate price behaviour during turbulent stock market periods. In particular, the market sentiment case affecting the bias parameter only exhibits a very good fit when compared to the empirical benchmark sample. When memory is added into the system, a strong impact around the BPD is observed. Memory also reinforces the ability to fit the benchmark pattern for particular setup combinations.
 	
To conclude, our approach suggests an alternative tool for examining the dynamics of changes in the HAM structure. To the best of our knowledge, this work also offers a first attempt to match the fields of HAM and behavioral finance in order to merge both approaches. We study the impact of `behavioural breaks' but other phenomena such as interest rate shocks or new supplies of information might be implemented as well. As such, our work might be viewed as a step toward an intriguing extension of HAM --- its empirical validation --- while we show that it makes sense to consider behavioural finance within the heterogeneous agent modelling framework and that both approaches can desirably complement one another.

%\section*{References}
\label{sec:ref}
\bibliography{refs}
\bibliographystyle{chicago}

\section*{Appendix}

\begin{sidewaystable}[!htbp]
\caption{DJIA Components 1987--2008}
\label{tab:stocks}
\centering
%\begin{tabular*}{1\textwidth}{@{\extracolsep{\fill}}llll} %pìkné zarovnání
%horizontal version: \begin{tabular}{p{3.5cm}p{2.6cm}p{5.8cm}p{0.6cm}}
\begin{tabular}{p{3.5cm}p{4.1cm}p{13.4cm}p{0.7cm}}
\hline
\noalign{\smallskip}
Event & \multicolumn{1}{c}{Period} & \multicolumn{1}{c}{Stocks (ticker symbols)} & \multicolumn{1}{c}{\#} \\
\hline
\noalign{\smallskip}
Black Monday & Sept 21 -- Oct 19, 1987 & AA, AXP, BA, CVX, DD, EK, FL, GE, GM, GT, HON, IBM, IP, KO, MCD, MMM, MO, MRK, NAV, PG, T, UTX, XOM; N/A: BS, PRI, `old' S, TX, UK, `old' WX, Xn & 460
\\
\noalign{\medskip}
   & Oct 19 -- Nov 16, 1987 & --$\vert\vert$-- & 460
\\
\hline
\noalign{\smallskip}
Ruble Devaluation & Jul 20 -- Aug 17, 1998 & AA, AXP, BA, C, CAT, CVX, DD, DIS, EK, GE, GM, GT, HON, HPQ, IBM, IP, JNJ, JPM, KO, MCD, MMM, MO, MRK, PG, T, UTX, WMT, XOM; N/A: `old' S, UK & 560
\\
\noalign{\medskip}
   & Aug 17 -- Sep 15, 1998 & --$\vert\vert$-- & 560
\\
\hline
\noalign{\smallskip}
Dot-com Bubble & Feb 10 -- Mar 10, 2000 & AA, AXP, BA, C, CAT, DD, DIS, EK, GE, GM, HD, HON, HPQ, IBM, INTC, IP, JNJ, JPM, KO, MCD, MMM, MO, MRK, MSFT, PG, T, UTX, WMT, XOM; N/A: IP only for Feb 23, 2000, SBC & 579
\\
\noalign{\medskip}
   & Mar 10 -- Apr 7, 2000 & --$\vert\vert$-- & 580
\\
\hline
\noalign{\smallskip}
WTC 9/11 Attack & Aug 10 -- Sep 10, 2001 & AA, AXP, BA, C, CAT, DD, DIS, EK, GE, GM, HD, HON, HPQ, IBM, INTC, IP, JNJ, JPM, KO, MCD, MMM, MO, MRK, MSFT, PG, T, UTX, WMT, XOM; N/A: SBC & 580
\\
\noalign{\medskip}
   & Sep 10 -- Oct 12, 2001 & --$\vert\vert$-- & 580
\\
\hline
\noalign{\smallskip}
Lehman Brothers & Aug 15 -- Sep 15, 2008 & AA, AXP, BA, BAC, C, CAT, CVX, DD, DIS, GE, GM, HD, HPQ, IBM, INTC, JNJ, JPM, KO, MCD, MMM, MRK, MSFT, PFE, PG, T, UTX, VZ, WMT, XOM; omitted as an outlier: AIG & 580
\\
\noalign{\medskip}
   & Sep 15 -- Oct 13, 2008 & --$\vert\vert$--, but omitted AIG was replaced by KFT on September 22, 2008 --- we omit KFT as well to keep dataset consistency & 580
\\
\hline
\multicolumn{4}{l}{\scriptsize{\emph{Note:} \# stands for number of observations. N/A stands for stock not available for a particular period.}}
\\
\multicolumn{4}{l}{\scriptsize{`old' denotes companies which historical ticker symbol is nowadays used for another company.}}
\\
\hline
    \end{tabular}
\begin{center}
\end{center}
\end{sidewaystable}

\end{document}